\documentclass[aip,jcp,reprint]{revtex4-1} 

\usepackage{subcaption}
\usepackage[font=small,labelfont=bf,justification=raggedright]{caption}
\usepackage[english]{babel}
\usepackage[version=4]{mhchem}
\usepackage{hyperref}
\usepackage{amsmath}
\usepackage{amsfonts}
\usepackage{amssymb}
\usepackage{graphicx} 
\usepackage{threeparttable}
\usepackage{booktabs}
\usepackage{siunitx}
\usepackage{svn-multi}
\usepackage{datetime}
\usepackage{textcomp}
\usepackage{multirow}
\usepackage{dblfloatfix}
\usepackage[colorinlistoftodos,color=green,textsize=tiny,textwidth=1.5cm]{todonotes}
\usepackage{nomencl} 
\usepackage{soul}
\usepackage{lipsum}
\usepackage{dcolumn}
\usepackage{adjustbox}

\usepackage[final]{changes}

\definechangesauthor[name={Reviewer 2. Comment 1}, color=black]{r21}
\definechangesauthor[name={Reviewer 2. Comment 2}, color=black]{r22}
\definechangesauthor[name={Reviewer 2. Comment 3}, color=black]{r23}
\definechangesauthor[name={Reviewer 1. Comment 1}, color=black]{r11}
\definechangesauthor[name={Reviewer 1. Comment 2}, color=black]{r12}
\definechangesauthor[name={Reviewer 1. Comment 3}, color=black]{r13}
\definechangesauthor[name={Reviewer 1. Comment 4}, color=black]{r14}
\definechangesauthor[name={Reviewer 1. Comment 5}, color=black]{r15}
\definechangesauthor[name={Reviewer 1. Comment 6}, color=black]{r16}
\definechangesauthor[name={Reviewer 1. Comment 7}, color=black]{r17}
\definechangesauthor[name={Reviewer 1. Comment 8}, color=black]{r18}
\definechangesauthor[name={Reviewer 1. Comment 9}, color=black]{r19}
\definechangesauthor[name={Reviewer 2. Comment 1 (2nd)}, color=red]{r21N}
\definechangesauthor[name={Reviewer 2. Comment 2 (2nd)}, color=red]{r22N}

\usepackage{lineno}  
\newcommand{\degree}{\ensuremath{^\circ}} 

\newcounter{todocounter}

\def\mys#1{{\mbox{\tiny{#1}}}}    

\def\dls{\collrad_{\mys{h}}}      
\def\collrad{R}                   
\def\ac{\collrad_{\mys{c}}}       
\def\poly{s}            		  
\def\CAOT{C_{\mys{AOT}}}          
\def\rhoI{\rho_{\mys{I}}}         
\def\cion{C_{\mys{ion}}}          
\def\shell{\delta}				  
\def\SM{S_{\mys{M}}}			  

\def\etac{\eta}        			  
\def\cross{\eta_{\mys{c}}}        
\def\olp{\eta^{*}}                

\def\volc{\eta_{\mys{c}}}		  
\def\rhoc{\rho_{\mys{C}}}         
\def\rp{\rho_{+}}                
\def\rhos{\rho_{\mys{s}}}         
\def\rm{\rho_{-}}                
\def\rsp{\bar{\rho}_{+}}         
\def\rsm{\bar{\rho}_{-}}         
\def\rs0{\bar{\rho}_{0}}         

\def\d{d}        				  
\def\h{h}        				  
\def\mured{\mu_{\mys{red}}}        
\def\visc{\nu}				      
\def\PhiS{\Phi_{\mys{s}}}	      
\def\kB{k_{\mys{B}}}  			
\def\kBT{\kB T}                 
\def\Na{N_{\mys{A}}}            
\def\debye{\lambda_{\mys{D}}}	
\def\lB{\ell_\mys{B}}    		
\def\kap{\kappa_{\mys{eff}}}	
\def\kres{\kappa_{\mys{res}}}	
\def\pot{U_{\mys{eff}}}	 					
\def\Z{Z_{\mys{eff}}}			
\def\bare{Z}			        
\def\RWS{R_{\mys{WS}}}			
\def\PhiD{\Phi_{\mys{D}}}	    
\def\leff{\lambda_{\mys{eff}}}	
\makenomenclature 

\begin{document}


\nomenclature[radiuscs]{$\collrad$}{Core-shell radius of colloidal particle \verb $\collrad$ }
\nomenclature[radiusc]{$\ac$}{Core radius of colloidal particle \verb $\ac$ }
\nomenclature[poly]{$\poly$}{Coefficient of radius variation (Polydispersity) \verb $\poly$ }
\nomenclature[CAOT]{$\CAOT$}{Molar concentration of AOT added \verb $\CAOT$ }          
\nomenclature[Ccmc]{$C_{\mys{CMC}}$}{Concentration of CMC \verb $C_{\mys{CMC}}$ }          
\nomenclature[cion]{$C_{\mys{ion}}$}{Total ion concentration \verb $\cion$ } 
\nomenclature[Cplus]{$C_{+}$}{Positive ion concentration \verb $C_{+}$ }
\nomenclature[Cminus]{$C_{-}$}{Negative ion concentration \verb $C_{-}$ }   
\nomenclature[dls]{$\dls$}{Hydrodynamic radius of particle \verb $\dls$ }
\nomenclature[d]{$\d$}{centre-to-centre spacing in suspension \verb $\d$ }
\nomenclature[eta]{$\etac$}{Colloid volume fraction \verb $\etac$ }
\nomenclature[etacross]{$\cross$}{Cross-over colloid volume fraction from salt to counterion screening regime \verb $\cross$ }
\nomenclature[etaolp]{$\olp$}{Colloid volume fraction at overlap of double layers \verb $\olp$ }
\nomenclature[volc]{$\volc$}{Core volume fraction \verb $\volc$ }
\nomenclature[epsilon0]{$\epsilon_{0}$}{Permittivity of a vacuum \verb $\epsilon_{0}$ }
\nomenclature[epsilonr]{$\epsilon_{\mys{r}}$}{Relative permittivity of solvent \verb $\epsilon_{\mys{r}}$ }
\nomenclature[e]{$e$}{Fundamental charge \verb $e$ }
\nomenclature[E]{$E$}{Electric field strength \verb $E$ }
\nomenclature[h]{$\h$}{Surface-to-surface separation \verb $\h$ }
\nomenclature[kap]{$\kap$}{Total inverse decay length in suspension \verb $\kap$ }
\nomenclature[lambda]{$\lambda$}{Mean particle spacing in units of $\kres^{-1}$ \verb $\lambda$ }
\nomenclature[lambdaeff]{$\leff$}{Mean particle spacing in units of $\kap^{-1}$ \verb $\leff$ }
\nomenclature[lambdaD]{$\debye$}{Debye screening length \verb $\debye$ }
\nomenclature[kres]{$\kres$}{Contribution to inverse decay length due to salt reservoir \verb $\kres$ }
\nomenclature[kB]{$\kB$}{Boltzmann constant \verb $\kB$ }
\nomenclature[kBT]{$\kBT$}{Thermal energy \verb $\kBT$ }
\nomenclature[lB]{$\lB$}{Bjerrum length \verb $\lB$ }
\nomenclature[Na]{$\Na$}{Avogadro's constant \verb $\Na$ }
\nomenclature[r]{$r$}{Radial variable \verb $r$ }
\nomenclature[rp]{$\rp$}{Positive ion number density \verb $\rp$ }
\nomenclature[rm]{$\rm$}{Negative ion number density \verb $\rm$ }
\nomenclature[rhoc]{$\rhoc$}{Particle number density \verb $\rhoc$ }
\nomenclature[rhos]{$\rhos$}{Univalent salt concentration \verb $\rhos$ }
\nomenclature[rho]{$\rho$}{Mass density \verb $\rho$ }
\nomenclature[rhoI]{$\rhoI$}{Total ion number density \verb $\rhoI$ }
\nomenclature[Z]{$\Z$}{Effective particle charge \verb $\Z$ }
\nomenclature[Zbare]{$\bare$}{Bare particle charge \verb $\bare$ }
\nomenclature[phi]{$\phi$}{Electrostatic potential \verb $\phi$ }
\nomenclature[Phi]{$\Phi$}{Dimensionless electrostatic potential \verb $\Phi$ }
\nomenclature[PhiS]{$\PhiS$}{Surface potential \verb $\PhiS$ }
\nomenclature[PhiD]{$\PhiD$}{scaled donnan potential \verb $\PhiD$ }
\nomenclature[visc]{$\visc$}{Solvent viscosity \verb $\visc$ }
\nomenclature[v]{$v$}{Particle velocity \verb $v$ }
\nomenclature[mured]{$\mured$}{Reduced mobility \verb $\mured$ }
\nomenclature[mured0]{$\mured^{0}$}{Reduced mobility as $\etac \rightarrow 0$ \verb $\mured^{0}$ }
\nomenclature[mu]{$\mu$}{Electrophoretic mobility \verb $\mu$ }
\nomenclature[n]{$n$}{Refractive index \verb $n$ }
\nomenclature[H]{$H$}{Integrand in electrophoretic calculations \verb $H$ }

\nomenclature[A]{$A$}{Instrument factor \verb $A$ }
\nomenclature[I]{$I$}{SAXS scattered intensity \verb $I$ }
\nomenclature[q]{$q$}{Wavevector \verb $q$ }
\nomenclature[f]{$f$}{Particle size distribution \verb $f$ }
\nomenclature[lambda0]{$\lambda_{0}$}{Wavelength of x-rays \verb $\lambda_{0}$ }
\nomenclature[theta]{$\theta$}{Scattering angle \verb $\theta$ }
\nomenclature[P]{$P$}{Normalized form factor \verb $P$ }
\nomenclature[delta]{$\shell$}{Shell thickness \verb $\shell$ }
\nomenclature[S]{$S$}{Structure factor (monodisperse) \verb $S$ }
\nomenclature[SM]{$\SM$}{Measured structure factor  \verb $\SM$ }

\nomenclature[chi]{$\chi$}{Conversion factor between molar AOT concentration and ion concentration \verb $\chi$ }
\nomenclature[alpha]{$\alpha$}{Scaled effective screening parameter \verb $\alpha$ = $ \kap \collrad $ }
\nomenclature[Delta]{$\Delta$}{Scaled charge from counterions \verb $\Delta$ }
\nomenclature[beta]{$\beta$}{Inverse thermal energy $= 1/ \kBT$ \verb $\beta$ }
\nomenclature[gamma]{$\gamma$}{Yukawa prefactor \verb $\gamma$ }
\nomenclature[x]{$x$}{Dimensionless radial variable $x = r / \collrad $ \verb $x$ }
\nomenclature[radiusWS]{$\RWS$}{Radius of WS cell  \verb $\RWS$ }
\nomenclature[U]{$\pot$}{Pair interaction potential \verb $\pot$ }
\nomenclature[T]{$T$}{Temperature \verb $T$ }
\nomenclature[K]{$K$}{AOT self-ionization equilibrium constant \verb $K$ }
\nomenclature[chiT]{$\chi_{\mys{T}}$}{Isothermal compressibility \verb $\chi_{\mys{T}}$ }
\nomenclature[Pi]{$\Pi$}{Osmotic pressure \verb $Pi$ }

\sisetup{separate-uncertainty}    

\graphicspath{{../figures/}}

\svnidlong
{$HeadURL: svn+ssh://it016641.chm.bris.ac.uk/home/cppb/svnprojects/SAXS-paper-2013/paper-2016/SAXS.tex $}
{$LastChangedDate: 2017-10-23 16:47:03 +0100 (Mon, 23 Oct 2017) $}
{$LastChangedRevision: 248 $}
{$LastChangedBy: cppb $}
\svnid{$Id: SAXS.tex 248 2017-10-23 15:47:03Z cppb $}

\title{Charge regulation of nonpolar colloids}
\author{James E. \surname{Hallett}}
\affiliation{School of Chemistry, University of Bristol, Bristol BS8 1TS, UK.}
\affiliation{School of Physics, University of Bristol, Bristol BS8 1TL, UK.}

\author{{David A.~J. Gillespie}}
\affiliation{School of Chemistry, University of Bristol, Bristol BS8 1TS, UK.}


\author{Robert M. \surname{Richardson}}
\affiliation{School of Physics, University of Bristol, Bristol BS8 1TL, UK.}

\author{Paul \surname{Bartlett}}
\email[Corresponding author. \\ \textit{E-mail address:} ]{p.bartlett@bristol.ac.uk}
\affiliation{School of Chemistry, University of Bristol, Bristol BS8 1TS, UK.}

\date{\today}



\begin{abstract}
	
Individual colloids often carry a charge as a result of the dissociation (or adsorption) of weakly-ionized surface groups.  The magnitude depends on the precise chemical environment surrounding a particle, which in a concentrated dispersion is a function of the colloid packing fraction $\etac$. Theoretical studies have suggested that the effective charge $\Z$ in regulated systems could, in general, \textit{decrease} with increasing $\etac$. We test this hypothesis for nonpolar dispersions by determining $\Z(\etac)$ over a wide range of packing fractions ($10^{-5} \le \etac \le 0.3$)  using a combination of small-angle X-ray scattering and electrophoretic mobility measurements.  \added[id=r21N]{All dispersions remain entirely in the fluid phase regime.} We find a complex dependence of the particle charge as a function of the packing fraction, with $\Z$ initially decreasing at low concentrations before finally increasing  at high $\etac$. We attribute the non-monotonic density dependence to a crossover from concentration-independent screening at low $\etac$, to a  high packing fraction regime in which counterions outnumber salt ions and electrostatic screening becomes $\etac$-dependent.  \added[id=r22N]{The efficiency of charge stabilization at high concentrations may explain the unusually high stability of concentrated nanoparticle dispersions which has  been reported\cite{Batista2015}.}

\end{abstract}


\maketitle 



\section{Introduction} \label{sec-intro}


Virtually all colloids carry a charge when immersed in an electrolyte.  The subsequent Coulombic interactions are crucial to a wide variety of processes including understanding  the programmed self-assembly of nanoparticles\cite{Batista2015}, the phase stability of  suspensions\cite{derjaguin:41,Verwey1999}, and the hierarchical architecture of virus structures\cite{2012}. A practical challenge is that the charge  is not fixed  \textit{a priori} but typically free to adjust through a chemical equilibrium\cite{Ninham1971}. Strongly acidic or basic groups tend to be fully dissociated, regardless of system parameters such as salt concentration or pH, while the dissociation of weak acid or base surface groups depends on the electrochemical potential\cite{Ninham1971}. As a result, colloids carrying weak ionic groups are often referred to as \textit{charge regulated}, in the sense that the effective surface-charge density is not fixed but adapts to minimize the free energy of the system with ions migrating on and off surface sites. To date, the concepts of charge regulation \textbf{(CR)} have been applied almost exclusively to aqueous systems. Examples include analysis of the electrostatic double-layer interactions between surfaces covered with protonated groups\cite{Ninham1971,Behrens1999b,Dan2002,Trefalt2016}, explanation of the extremely long-range  attractive forces that operate between proteins with dissociable amino acid groups close to their point of zero charge\cite{Kirkwood1952,Lund2013,Adzic2015}, or \replaced[id=r11]{the role of charge regulation effects in determining not only the magnitude but the sign of the force between asymmetrically-charged particles\cite{Popa2010,Cao2017}.}{prediction of the strength of anisotropic repulsions between charge regulating ligands on nanoparticles.} In contrast, little is known regarding the role of CR in  nonpolar systems, where the degree of dissociation of surface ionic groups is small yet finite and complex charging processes operate\cite{Lee2016}. Indeed the strong long-range repulsive interactions arising from Coulombic charges in low dielectric solvents make nonpolar systems a fascinating testing ground for CR concepts. 


The double-layer interactions between charge-regulating particles are different to those between fully dissociated particles\cite{Everts2016,Smallenburg2011}. For example, from Poisson-Boltzmann theory the repulsive disjoining pressure $\Pi(\h)$ between identically-charged surfaces spaced by $\h$, in the limit of small separations ($\h \rightarrow 0$),  is fixed directly by the osmotic pressure of the ions which are trapped in the gap between the two surfaces by  electroneutrality constraints\cite{Andelman1995}. In a CR system the degree of dissociation of surface groups and hence the concentration of released counterions is controlled by the total electrostatic potential $\Phi (\h)$, which is itself a function of the separation $\h$ between the charged surfaces. As a result, the effective surface charge density and $\Pi$ differ substantially from the predictions of a constant-charge (\textbf{CC}) model as the two charge-regulated surfaces are brought together. In general, it has been found that CR significantly weakens the repulsive interactions between surfaces\cite{Markovich2016,Behrens1999b,Ninham1971}, as the increase in counterion osmotic pressure is relieved by a shift in the dissociation equilibria towards the uncharged state. 

The effect of charge regulation on the repulsive interactions between particles in concentrated dispersions has been examined in a number of studies\cite{Everts2016,Smallenburg2011}.  To simplify the many-body electrostatic problem, a Poisson-Boltzmann (PB) cell approximation has been frequently employed in which the dispersion is divided up into identical spherical cells, each containing just one colloid in osmotic equilibrium with a salt reservoir of Debye length $\kres^{-1}$. Cell calculations using different CR schemes\cite{Everts2016,Smallenburg2011} show that, independent of the specific surface chemistry, the colloidal particle \textit{discharges monotonically with increasing packing fraction $\etac$} -- in the sense that the colloid charge $\Z$ reduces with  increasing $\etac$. This is in line with the asymptotic dependence observed for charge-regulated plates as $\h \rightarrow 0$. The charge reduction is predicted to be most severe in dispersions of small particles and solutions of low ionic strength where $\kres \collrad \rightarrow 0$. 

%
%
%

The experimental situation is however less clear, which may be accounted in part by the focus so far on systems with relative large $\kres \collrad$. \citet{Royall2006}, using charges estimated from radial distribution functions measured at $\kres \collrad \approx 1$, have argued that the sequence of reentrant transitions observed upon increasing the colloid density in some charged nonpolar suspensions (fluid $\leftrightharpoons$ BCC $\leftrightharpoons$ fluid $\leftrightharpoons$ FCC), is a consequence of a steady \textit{reduction} in $\Z$ with increasing $\etac$. \added[id=r21]{In recent work,  \citet{Kanai2015} have explored the crystallization of large ($\collrad > \SI{0.46}{\micro \meter}$) colloids, charged by the addition of the surfactant Aerosol-OT (sodium bis(2-ethyl 1-hexyl)sulfosuccinate or AOT) in a nonpolar solvent mixture. They observed close agreement between the reentrant phase boundaries measured and numerical calculations of the electrostatic charging effects produced by the reverse micelles.}  In an alternative approach, \citet{Vissers2011} measured the electrophoretic mobility of concentrated nonpolar dispersions at $\kres \collrad \approx 0.5$.  Using an approximate cell model to take into account double-layer overlap, they showed that the particle charge reduced with increasing concentration. 

If the strength of the electrostatic repulsions between charge-regulated particles decays rapidly with increasing colloid packing fraction, it is feasible that concentrated dispersions of charged nanoparticles could become colloidally unstable at very low ionic strengths, due to the absence of a strong enough osmotic repulsion to counter attractive van der Waals forces\cite{Lin2012}.  Indeed, the notion that a high colloid surface-potential (and consequently a large particle charge) at low colloid concentrations does not necessarily guarantee stability has been proposed for some time. Over fifty years ago, \citet{Albers1959,albers_stability_1959}  noted the absence of any correlation between the electrokinetic $\zeta$-potential of water-in-benzene emulsions and their stability against coalescence. Later work by \citet{mishchuk_interparticle_2004} suggested that a charge-stabilized water-in-oil emulsion should be unstable above a critical volume fraction, which reduced as the ion concentration fell. Indeed there have been repeated experimental reports\cite{bensley_coagulation_1983,van_mil_stability_1984,green_stability_1988,wang_surface_2013} of rapid  coagulation at high colloid concentrations and low electrolyte concentrations in dispersions with high surface potentials, which would be sufficient to stabilize a more dilute system. 

In this article, we present a study of the packing fraction dependence of the effective charge and the bulk correlations in nonpolar colloids in the weak screening regime ($\kres \collrad \ll 1$). Experiments were carried out over an extended range of concentrations ($10^{-5} \le \etac \le 0.3$) using small nanoparticles in solutions of very low ionic strengths, so the dimensionless screening parameters $\kres \collrad$ studied are considerably smaller than in any previous work. This has the advantage that the charge reduction due to regulation is evident at significantly lower packing fractions. By using two different methods to prepare charged nonpolar dispersions, we systematically varied the background ion concentration within the range $0.0 \le \kres \collrad \le 0.24$. We determined the effective charge $\Z(\etac) $ as a function of $\etac$ by an analysis of (i) the particle structure factor obtained from X-ray scattering measurements, and (ii) the density dependence of the electrophoretic mobility at $\etac < 10^{-2}$. At low packing fractions, we observe the charge reduction predicted by theory for a charge-regulated system at low $\kres \collrad$. However at high particle concentrations ($\etac > 10^{-2}$) we demonstrate a surprising density dependence of the effective charge, with $\Z$ \textit{increasing} with $\etac$. Extensive measurements reveal that this is a general feature of concentrated charged dispersions at low ionic strengths. \added[id=r21N]{Our dispersions always remain entirely in the fluid regime so the changes identified are not a consequence of phase changes, such as crystallization.}  Our paper is organized as follows: In Sec.~\ref{sec:cell} existing CR models are summarized and we show that, although details differ, all models predict a monotonic reduction in the effective charge with increasing packing fraction. Details of the nanoparticle systems and the analysis methods used in our experiments are detailed in Sec.~\ref{sec-exp}. In Sec~\ref{sec:characterization} we confirm that, without added charge, our colloids display pure hard-sphere interactions. The electrophoretic mobility of dilute charged dispersions is analysed in Sec.~\ref{sec:mobility} and shown to be consistent with a simple CR model. \deleted[id=r22N]{The non-monotonic behaviour of} \replaced[id=r22N]{The }{ the} charge at high packing fractions is extracted from measured structure factors in Sec.~\ref{sec:high-conc}.  \added[id=r22N]{Interestingly, we find that the measured charge \textit{increases}, rather than decreasing with particle concentration as would be expected naively. This observation suggests that charge stabilization is much more effective in highly concentrated suspensions than is generally believed.} We discuss the origins of the surprising increase in the particle charge with concentration in Sec.~\ref{sec:charge-origin}, before concluding in Sec~\ref{sec:conclude}.

\section{Charge regulation models} \label{sec:cell}

We consider dispersions of $N$ charged nanoparticles of radius $\collrad$ and charge $\bare e$ suspended in a solvent of volume $V$, and relative permittivity $\epsilon_{\mys{r}}$ at a temperature $T$. The colloid number density is $\rhoc = N/V$ and the corresponding packing fraction is $\etac = (4 \pi/3) \rhoc \collrad^{3}$. The double-layer repulsions between particles are controlled by two length scales: the solvent-specific Bjerrum length $\lB = \beta e^{2} / (4 \pi \epsilon_{\mys{0}} \epsilon_{\mys{r}})$ where $\beta = 1/\kBT$, $\kB$ is the Boltzmann constant, $e$ is the fundamental charge, and $\epsilon_{\mys{0}}$ is the permittivity of a vacuum, and the (reservoir) Debye screening length $1/\kres = 1/\sqrt{8\pi \lB \rhos }$, with  $2\rhos$  the number density of univalent salt ions in the reservoir. In dodecane at $T=\SI{293}{\kelvin}$, the Bjerrum length is \SI{28.3}{\nano \meter}.

Although at high packing fractions or for thick double layers many-body interactions between charged nanoparticles can become significant\cite{Klein2002,Russ2002}, ab-initio computer simulations\cite{Loewen1993,Dobnikar2006} show that the colloidal structure  of a concentrated dispersion can often be approximated remarkably well by representing the effective pair potential $\pot$ by the hard-sphere Yukawa (HSY) function,
\begin{equation}\label{eq:Yukawa}
\beta \pot (x) = 
\begin{cases} 
\gamma \frac{e^{- \alpha x }}{x}, &  \text{for } {x \geq 2}  \\
\infty, &   {x < 2} \\
\end{cases}
\end{equation}
where $x = r / \collrad$ is the dimensionless pair separation, $r$ is the centre-to-centre distance, and the range and strength of the interparticle repulsions are evaluated from the effective charge $\Z$ and the screening length $\kap^{-1}$, according to 
\begin{equation}\label{eq-alpha}
\alpha = \kap \collrad
\end{equation}
and
\begin{equation}\label{eq:gamma}
\gamma  = \left(\frac{\Z \lB}{\collrad} \right)^{2}  \frac{\collrad}{\lB} \left (\frac{e^{\alpha}}{1+\alpha} \right)^{2}.
\end{equation}
While the HSY potential has been widely used, there is no rigorous route to construct such a potential at a finite nanoparticle density. In the low density limit $\etac \rightarrow 0$, Eq.~\ref{eq:Yukawa}  does reduce, under Debye-Huckel conditions ($|\Z| \lB / \collrad < 1$), to the classical expressions derived by Derjaguin and Landau\cite{derjaguin:41} and Verwey and Overbeek\cite{Verwey1999} with $\Z = \bare$ and $\kap = \kres$. In general, however the effective parameters $\Z$ and $\kap$ will be density dependent. Denton, for instance, has derived\cite{Denton2000} an expression similar to  Eq.~\ref{eq:Yukawa} from a  rigorous statistical mechanical treatment  but with a modified screening parameter that incorporates both the counterions released from the particle and corrects for excluded volume effects. For a colloid of charge $\Z$, and added 1:1 electrolyte of density $\rhos$ the screening parameter is obtained as
\begin{equation}\label{eq:Denton}
\kap =  \sqrt{ 4 \pi \lB\left( 2\rhos + \rhoc |\Z| \right)}
\end{equation}
where $\rhoc$ is the colloid number density. In this work, we follow this approach and fix the effective screening parameter $\kap$ using Eq.~\ref{eq:Denton}.

To illustrate the consequences of charge regulation, we determine the  effective charge $\Z$ using a PB-cell model of the dispersion, in which each particle  is placed at the centre of a Wigner-Seitz spherical cell of radius $\RWS = \collrad \etac^{-1/3}$. The electrostatic potential $\phi(r)$  is then a spherically-symmetric function of the radial variable $r$, measured from the centre of cell. To facilitate a direct comparison between theoretical predictions and experiment we compute the  scaled electrostatic potential $\Phi = \beta e \phi$ and the reduced effective charge $\Z \lB / \collrad$ as a function of $\etac$ by solving numerically the non-linear Poisson-Boltzmann equation on the interval $1 \le x \le \etac^{-1/3}$,
\begin{equation}\label{eq:PB}
2 \Phi'(x) + x\Phi''(x) = (\kres \collrad)^{2} x  \sinh \Phi(x)
\end{equation}
where $x = r/\collrad$.  Equation~\ref{eq:PB} is solved subject to the appropriate charge regulation boundary conditions\cite{Smallenburg2011,Everts2016}  together with the constraint,  $\Phi' (\etac^{-1/3}) = 0$, which follows from the overall charge neutrality of the Wigner-Seitz cell. The effective charge $\Z$ is obtained from the numerically-obtained electrostatic potential by matching $\Phi(x)$ at the edge of the cell to a solution of the linearized PB equation\cite{Trizac2003}.

\begin{figure}[h]
  \begin{subfigure}{0.85\linewidth}
       \includegraphics[width=\textwidth]{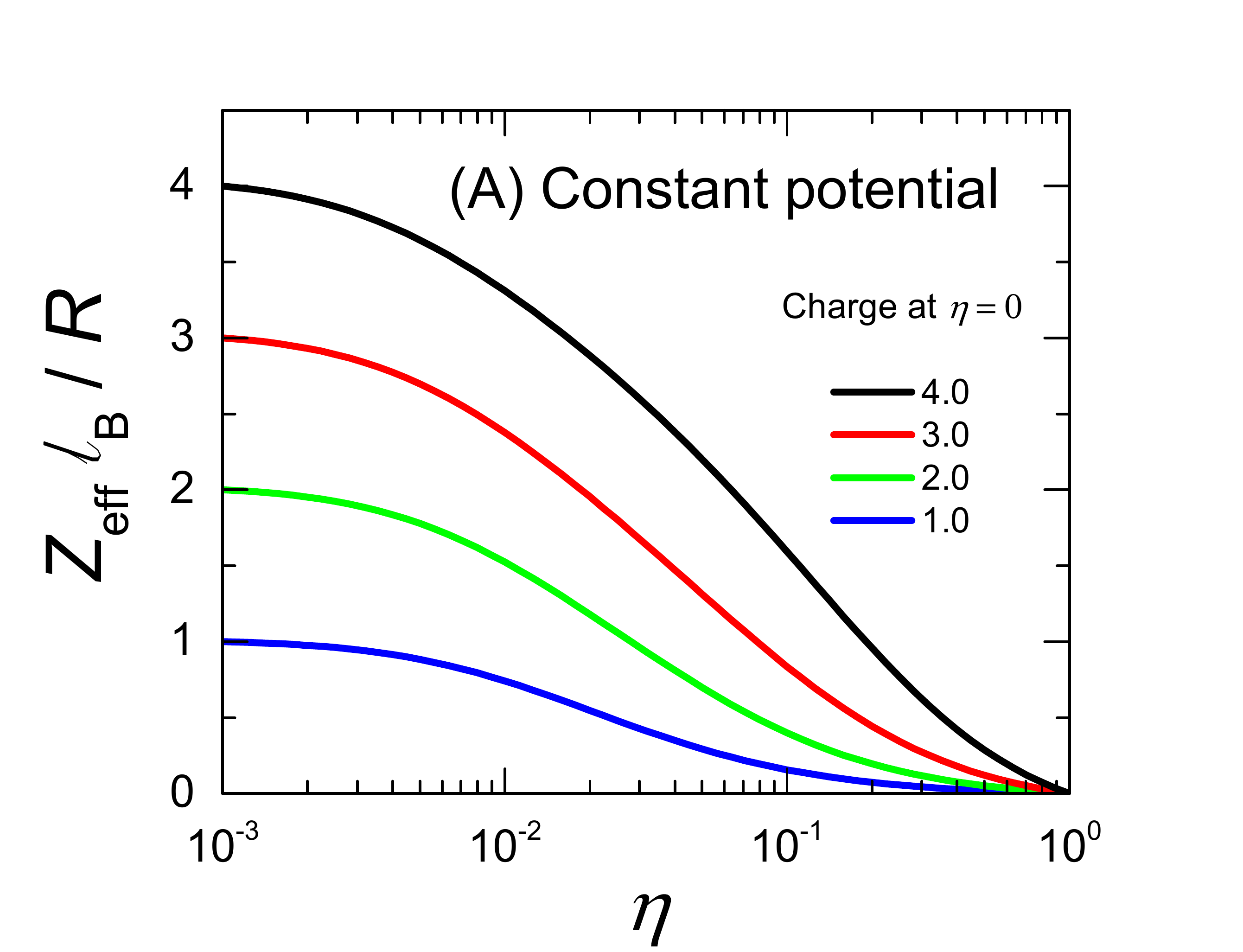}
  \end{subfigure}
  \begin{subfigure}{0.85\linewidth}
        \includegraphics[width=\textwidth]{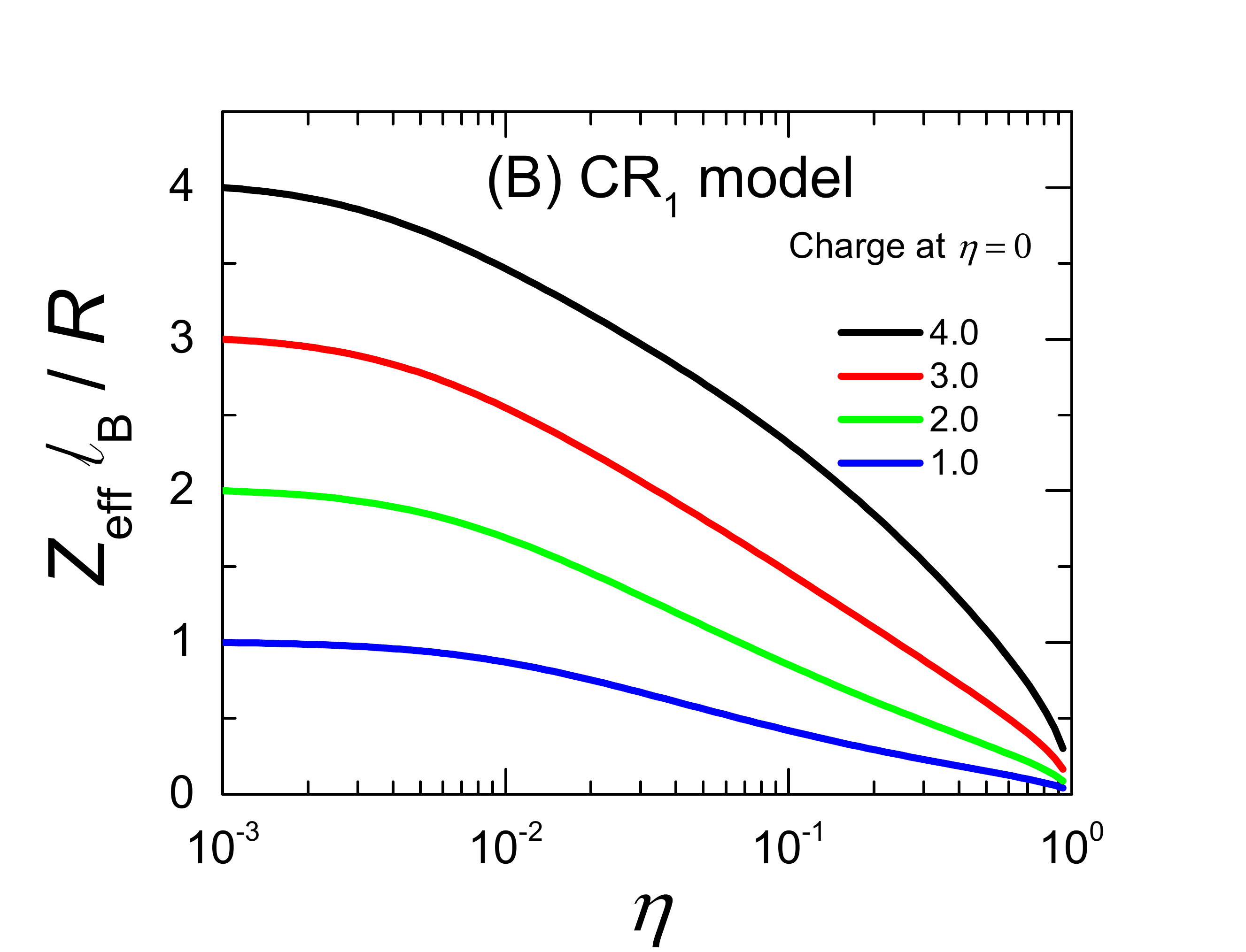}
  \end{subfigure}
  	\caption{Packing fraction dependence of the reduced charge $\Z \lB / \collrad$ predicted by charge regulation models for $\kres \collrad = 0.25$. (A) Constant potential (CP), and (B) single site association/dissociation  model (CR$_{1}$). Model parameters are chosen such that, in both cases, the reduced charge is fixed in the dilute limit. \deleted[id=r17]{, at $ \Z \lB / \collrad =[1\cdots 4]$}}
  	\label{fgr:extrapolated-point-charges}
  \end{figure}

The simplest theoretical model of a charged colloid is as a spherical insulator with a \textit{constant charge}  (\textbf{CC model}). The CC model implies no exchange between surface binding sites and free ions in solution so that the bare  charge remains frozen, independent of the electrolyte concentration and the particle packing fraction $\etac$. 
Recently, there have been a number of attempts\cite{Roberts2008,Smallenburg2011,Everts2016} to develop more realistic statistical models of charge regulation in a low dielectric environment. Fig.~\ref{fgr:extrapolated-point-charges} shows numerical predictions for the dimensionless charge
$\Z \lB / \collrad$ as a function of the packing fraction $\etac$, for a number of different surface chemistries. Crucially, we see that while the details of each model differ, they share a common qualitative behavior in that the particles are predicted to discharge continuously with increased colloid density.

 \citet{Everts2016}, for instance, have proposed that the  net colloid charge  is  determined by a  balance between two \textit{competing} surface ionization reactions,
\begin{eqnarray} \label{eq-CR2}
\ce{S1 + P+ <--> S1P+}, \nonumber \\
\ce{S2 + N- <--> S2N-},
\end{eqnarray}
 where S$_{1,2}$ denote different sites on the colloidal surface which  bind either positive (\ce{P+}) or negative ions (\ce{N-}). We label the generic two-site charge regulation scheme, outlined in Eq.~\ref{eq-CR2}, as an example of a \textbf{CR$_{2}$ model}. Numerical solution of this model reveals  that the charge $\Z$ carried by the colloidal particle reduces monotonically with increasing colloid density, in contrast to the CC model. Effectively, the particle  discharges continuously as a function of $\etac$, with the charge asymptotically tending to $\Z \approx 0$ as $\etac \rightarrow 1$. 
 
 In the limit, where there is  a significant adsorption of \textit{both} positive and negative ions onto the particle, the CR$_{2}$ boundary condition reduces to the simpler constant potential (CP) model.
  The  ionization of the positive and negative surface groups adjusts  so that as particles approach each other counterions migrate back onto surface sites to maintain a fixed surface potential, and the surface charge density decreases monotonically with increasing colloid density. Theoretical arguments for the validity of a CP model in nonpolar dispersions  have been made by a number of  authors\cite{Hsu2005,Roberts2008,Smallenburg2011,Everts2016}. \citet{Roberts2008} have, for instance, analysed a model of charge regulation in which charged micelles adsorb onto the surface of a colloid and shown that it is equivalent to assuming constant potential boundary conditions provided that (a) both positive and negative micelles are able to adsorb and (b) the surface coverage of  micelles is below the saturation limit. 
  
  Finally, \citet{Smallenburg2011} has compared numerical predictions for $\Z$ as a function of $\etac$ from a single-site association/dissociation equilibrium (\textbf{CR$_{1}$ model}) 
 \begin{equation}\label{eq+CR1}
 \ce{S + P+ <--> SP+},
 \end{equation}
 with CP calculations and found that although the predicted surface charges are indistinguishable at low densities, there are significant variations at higher $\etac$. These observations are consistent with  more general predictions\cite{Markovich2016} that the repulsions between surfaces with both positive and negative sites more closely resemble the constant potential limit than a surface with only a single ionizable site. The CR$_{1}$ scheme however still predicts a qualitatively similar charge dependence to the CR$_{2}$ model, in  that $\Z$ decreases monotonically from a finite low-$\etac$ value to essentially 0 at $\etac \approx 1$.

\section{Materials and methods} \label{sec-exp}

\subsection{Colloids} \label{sec:nano}

The systems studied consisted of three batches of \replaced[id=r12]{nanoparticles }{particles} (NP1-NP3)\added[id=r12]{, approximately \SI{50}{\nano \meter} in radii,} dispersed in dry dodecane (dielectric constant $\epsilon_{\mys{r}} = 2.01$) at a packing fraction $\etac$.    Each system consisted of a core of poly(methyl methacrylate) [PMMA]  surrounded by a chemically-grafted shell of poly(12-hydroxystearic acid) [PHSA].   The particles were prepared in-house\cite{Antl1986,Hussain2013} by a free-radical dispersion polymerization of methyl methacrylate (MMA) and methacrylic acid (MAA) in a mass ratio of 98:2 using 2,2\textasciiacute-azobis(2-methylpropionitrile) [AIBN] as initiator and a preformed graft copolymer poly(12-hydroxystearic acid)-\textit{g}-poly(methyl methacrylate) as dispersant. The synthesis was carried out at \SI{80}{\celsius} in a mixed  solvent of dodecane and hexane (2:1 by wt.) for \SI{2}{\hour}, before the temperature was raised to \SI{120}{\celsius} for a further \SI{12}{\hour},  in the presence of a catalyst, to  covalently lock the stabilizer to the surface of the particle.  All particles were undyed. Colloids were purified by repeated cycles of centrifugation and redispersion in freshly-dried dodecane to remove excess electrolyte and stabilizer.  \added[id=r15]{Once cleaned, the average hydrodynamic radius $\dls$ was determined by dynamic light scattering (DLS) using a Malvern Zetasizer nano S90 (Malvern instruments, UK). The sizes are given in Table~\ref{tblparticle}.}  \replaced[id=r15]{Nanoparticles} {The particles } were stored under nitrogen to prevent water uptake and 	ion generation.  The conductivity of the purified dispersions  \replaced[id=r15]{was }{were} checked prior to use with a Scientifica (UK) model 627 conductivity meter.


We employed two routes to generate charge: batches NP1 and NP2 were \added[id=r13]{ differently-sized PMMA particles, with the same surface chemistry, which were both} charged negative\cite{Hsu2005,Roberts2008} by the addition of the surfactant \replaced[id=r21]{AOT}{Aerosol-OT [AOT, sodium bis(2-ethyl 1-hexyl)sulfosuccinate]} while \added[id=r13]{nanoparticles} NP3 \added[id=r13]{had a different surface chemistry, and were} charged positive by the dissociation of lipophilic ionic groups\cite{Gillespie2013,Hussain2013}  introduced into the particle during synthesis.
AOT (98\%, Aldrich, UK) was purified by dissolution in dry methanol and centrifuged prior to use to remove residual salts. It was  used at molar concentrations $\CAOT$ above the critical micellar concentration\cite{Smith2013} $C_{\mys{CMC}} \approx $ \SI{0.13}{\milli \mole \per \deci \meter \cubed }, where although the majority of the reverse micelles are neutral conductivity measurements\cite{Eicke1989,Roberts2008} show that a small fraction of the micelles ($\approx 1$ in $10^{5}$) are ionized by thermal fluctuations.  Measurement of the conductivity and viscosity of AOT in dodecane confirmed that the total molar concentration of reverse micellar ions $\cion = C_{+} + C_{-}$ increased linearly\cite{Roberts2008} with $\CAOT$, $\cion = \chi \CAOT$ with $\chi = 5.5 \times 10^{-7}$.   A range of AOT concentrations, from $\CAOT = 5-225$ \si{\milli \mole \per \deci \meter \cubed} was employed, which correspond to total micellar ion concentrations of $2.8-125$ \si{\nano \mole \per \deci \meter \cubed}.  Suspensions were prepared by dilution of the same concentrated particle stock for all AOT concentrations to ensure accurate relative concentrations.  The dominant ionic species are  singly-charged positive  and negative AOT micelles. The colloid charge is not fixed but regulated by a competitive adsorption of cationic and anionic micellar ions onto the particle surface\cite{Roberts2008}.
Batch NP3, in contrast, contains no-added salt and is a counterion-only system. Particles were charged by copolymerization into the core of the particle of approximately 4 wt\% of the ionic monomer  \textit{n}-tridodecyl-propyl-3-methacryloyloxy ammonium tetrakis [3,5-bis (trifluoromethyl) phenyl] borate (\ce{[ILM-(C12)]+}\ce{[TFPB]-}). \added[id=r14]{The molecular structure of the polymeric NP3 nanoparticles is described, in greater detail, in the supplementary information.} The ionic monomer and nanoparticles were prepared following the procedures outlined in previous work\cite{Gillespie2013,Hussain2013}. Ionic dissociation of surface-bound \ce{[ILM-(C12)]+}\ce{[TFPB]-} groups generated a positive colloid charge of $+Z$ together with $Z$ negative \ce{[TFPB]-} counterions in solution\cite{Gillespie2013}.

\subsection{Electrophoretic mobility}

The electrophoretic mobility $\mu$  as a function of packing fraction $\etac$ was measured at \SI{25}{\degreeCelsius} using  phase-analysis light scattering (Zetasizer Nano, Malvern, UK).  The mobility $\mu = v / E$, where $v$ is the electrophoretic velocity induced by an applied electric field of strength $E$, was determined from the modulation in the phase of scattered light produced by a periodic triangular $E$-field. Equilibrated samples  were transferred into clean \SI{10}{\milli \meter} square quartz-glass cuvettes and a non-aqueous dip cell (PCS1115, Malvern) with a \SI{2}{\milli \meter} electrode gap placed in the cell. In a typical measurement, a series of runs were performed at different driving voltages between 10 V and 50 V and the scattering from a \SI{633}{\nano \meter} laser was collected at a scattering angle of \SI{173}{\degree}. No systematic dependence of $\mu$ on $E$ was observed. Any measurement where the phase plot significantly deviated from the expected triangular form was discarded,  and the measurement repeated. Since PMMA nanoparticles  in dodecane are weakly scatterers of light, reliable electrophoretic mobilities were recorded over a relatively wide range of packing fractions, $10^{-5} \le \etac \le 10^{-2}$. 

\subsection{Modelling of electrophoretic mobility}\label{sec:vol-fract-mu}

The electrophoretic mobility $\mu$ of a colloidal particle is determined by a balance between electrostatic and hydrodynamic forces. In the limits of low concentration and $\kres \collrad \rightarrow 0$ the reduced mobility, defined by the expression $\mured = 6 \pi \visc \lB \mu / e$ (where $\visc$ is the solvent of viscosity), is equal to the scaled particle charge $\Z \lB / \collrad$.
%
At finite concentrations however, the mobility decreases approximately logarithmically\cite{Lobaskin2007} with increasing $\etac$, as frictional forces grow because of strengthening particle-particle interactions.  To model the effects of interactions, we follow the theoretical analysis of Levine and Neal\cite{Levine1974} who  proposed a cell model for the mobility in a concentrated dispersion, valid in the linear PB limit.   Ohshima\cite{Ohshima1997a} derived an equivalent expression, 
\begin{eqnarray}\label{eq:mob1}
\mured & = &  \int_{1}^{\etac^{-1/3}} H(x)\left (1+ \frac{1}{2x^{3}} \right) \mathrm{d}x \nonumber \\
& + & \frac{\PhiD}{3} \frac{(\kres \collrad)^{2}}{1-\etac} \left(1+\frac{\etac}{2} \right) \left (1+\frac{1}{\etac}-\frac{9}{5 \etac^{2/3}}-\frac{\etac}{5} \right)
\end{eqnarray}
where  $\PhiD = \Phi(\etac^{-1/3})$ is the potential at the cell boundary  and, the integrand $H(x)$ is a function of the scaled potential  within the cell,
\begin{eqnarray}\label{eq:mob2}
H(x) & = &  -\frac{(\kres \collrad)^{2}}{6(1-\etac)}\left [1-3x^{2}+2x^{3} \cdots \right. \nonumber \\
& & \left.  -\etac\left( \frac{2}{5} - x^{3} + \frac{3}{5} x^{5}\right)  \right] \Phi'(x)
\end{eqnarray}
with $x = r/\collrad$.

\subsection{Small-angle X-ray scattering (SAXS)}\label{sec-SAXS}

SAXS measurements were performed at a temperature of \SI{20}{\degreeCelsius} on the Diamond Light Source (Didcot, UK) using the I22 beamline at a wavelength of $\lambda_{0} = $ \SI{0.124}{\nano \meter} and a sample to detector distance of \SI{10}{\meter} leading to a useful $q$-range of approximately 0.015 --  \SI{0.7}{\per \nano \meter }.  The scattering wave vector $q$ is defined as
\begin{equation}\label{eqq}
q = \frac{4 \pi}{\lambda_{0}} \sin (\theta /2)
\end{equation}
where $\theta$ is the scattering angle and $\lambda_{0}$ is the incident X-ray wavelength. Dispersions were loaded into reusable flow-through quartz capillary cells that were filled alternately with samples of the background solvent and the dispersion to allow accurate subtractions of the background. At least ten 2D images of \SI{10}{\second} each were collected, azimuthally averaged, transmission and background corrected according to established procedures to yield the scattered intensity $I(q)$, as a function of $q$. An additional series of higher resolution measurements were made at the ESRF (Grenoble, France) on the ID02 beamline with a $q$-range of approximately 0.01 --  \SI{0.8}{\per \nano \meter }.

The  intensity scattered by a dispersion of spherical particles with a narrow size distribution $f(\collrad)$ can be factored as 
\begin{equation}\label{key}
I(q) = A \etac P_{\mys{M}}(q) \SM(q), 
\end{equation}
where $A$ is an instrumental factor, $\etac$ is the packing fraction, $P_{\mys{M}} (q) = \int P(q;\collrad) f(\collrad) \mathrm{d}\collrad$ is the polydisperse particle form factor, and $\SM(q)$ is the measured structure factor\cite{Narayanan2007}. This was determined experimentally  from the intensity ratio, 
\begin{equation}\label{key}
\SM(q) = (\etac_{\mys{dil}} / \etac) \frac{I(q)}{I_{\mys{dil}}(q)},
\end{equation} 
where the dilute scattered intensity $I_{\mys{dil}}(q) = A \etac_{\mys{dil}} P_{\mys{M}} (q)$ was measured at a sufficiently low packing fraction ($\etac_{\mys{dil}} \approx 10^{-3}$)  to ensure that all interparticle interactions were suppressed and $\SM(q) = 1$.

\subsection{Charge screening} \label{sec:mat-kappa}

 The effective screening length in the micelle-containing systems can not be calculated directly from Eq.~\ref{eq:Denton} because the micellar ions are in equilibrium with neutral micelles, through an auto-ionization reaction of the form  \ce{M+ + M- <=> 2M}. The mixture of charged and neutral micelles acts effectively as a charge buffer.
To model the buffering process, we characterize the self-ionization equilibrium constant as $K = \rp \rm$, with $\rho_{\pm}$  the number density of the $\pm$ micelles. If the colloid has a surface charge of $-\Z$, then, from charge neutrality, the number density of positive ions in solution is $\rp = \rm + \rhoc \Z$. Adopting the particle radius $\collrad$ as a natural length scale and introducing the scaled ion densities $\bar{\rho}_{\pm} = \rho_{\pm} \collrad^{3}$ then we may express the scaled positive ion density simply as $\rsp = \rsm + 2 \Delta$, where $\Delta = 3 \etac \Z / (8 \pi)$.

Substituting this expression into the law of mass action gives the equilibrium ion concentrations, in the presence of charged nanoparticles, as
\begin{align}
\rsp &=  \sqrt{\Delta^{2}+(K\collrad^{3})^{2}} + \Delta \label{eq:buffer} \nonumber \\ 
\rsm &= \sqrt{\Delta^{2}+(K\collrad^{3})^{2}} -\Delta
\end{align}
The total ion concentration is therefore $\rsm + \rsm = 2  \sqrt{\Delta^{2}+(K\collrad^{3})^{2}}$, which is less than the value of $2(\Delta + K\collrad^{3})$ obtained by naively adding the salt and counterion densities together, demonstrating the charge buffering effect. From Eq.~\ref{eq:buffer} the corresponding screening  parameter is 
\begin{equation}\label{NP1,NP2}
\left (\kap \collrad \right )^{2}  = 8 \pi \left ( \frac{\lB}{\collrad} \right ) \sqrt{\Delta^{2}+(K\collrad^{3})^{2}} 
\end{equation}
which in the limit of no background ions ($K =0$) reduces to the counterion-only limit
\begin{equation}\label{NP3}
\left (\kap \collrad \right )^{2}  =  3 \etac \left ( \frac{|\Z| \lB}{\collrad} \right).
\end{equation}

%

%

\section{Results and Discussion}

\subsection{Nanoparticles}\label{sec:characterization}

We used  nanoparticles with a core of poly(methyl methacrylate) [PMMA] of radius $\ac$ sterically stabilized by a chemically-grafted shell of poly-12-hydroxystearic acid [PHSA] of thickness $\shell$ suspended in dodecane. To determine \replaced[id=r16]{the core radius $\ac$,}{these parameters,} the excess (nanoparticle dispersion minus solvent) SAXS scattering profiles $I(q)$ were measured from dilute dispersions ($\etac \approx 10^{-3}$). The X-ray scattering length density of the core was calculated to be \SI{10.8 e-6 }{\per \angstrom \squared} and the stabiliser layer as \SI{8 \pm 1 e-6 }{\per \angstrom \squared}, where the uncertainty is due to the variation in mass density reported in the literature. The calculated X-ray scattering length
density of dodecane is  \SI{7.34 e-6 }{\per \angstrom \squared}, so the  shell contrast is weak and the scattering arises predominately from the PMMA core. To model the dilute particle data, we used a polydisperse core-shell model with a Schulz size distribution\cite{DiCola2009} adjusting the (mean) core radius $\ac$ and the polydispersity $\poly$ to best describe the measured $I(q)$. The shell thickness was fixed at $\shell = $ \SI{10}{\nano \meter} on the basis of previous measurements\cite{Liddle2011,DiCola2009}. Agreement between the model calculations and the low-$\etac$ SAXS data is very good, with the fitted values of $\ac$ and $\poly$ listed in Table~\ref{tblparticle}.

\begin{figure}[tbp]
    %
    %
    %
	%
    \begin{subfigure}[c]{0.85\linewidth}
        \includegraphics[width=\textwidth]{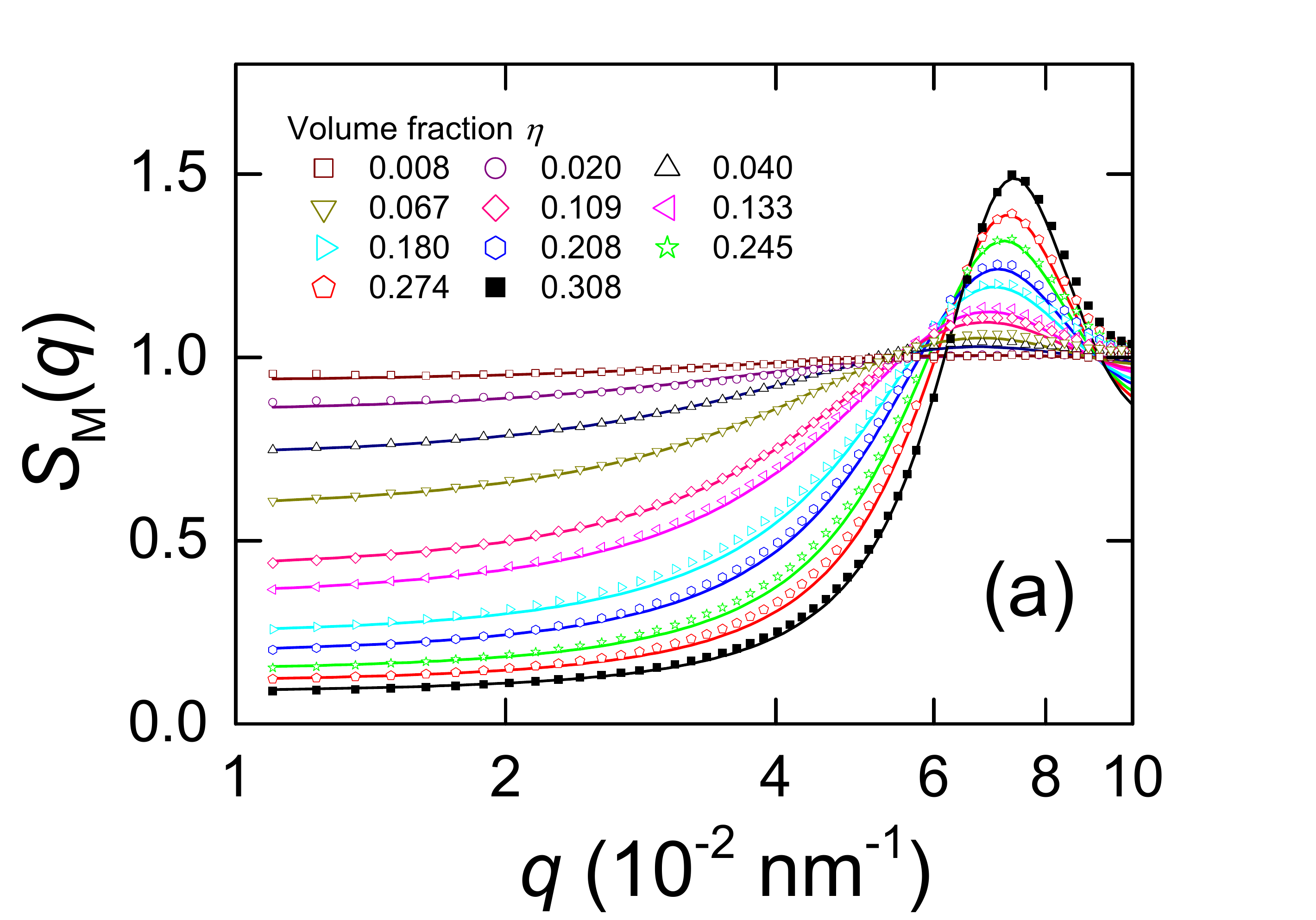}
    \end{subfigure}
    \begin{subfigure}[c]{0.85\linewidth}
		\includegraphics[width=\textwidth]{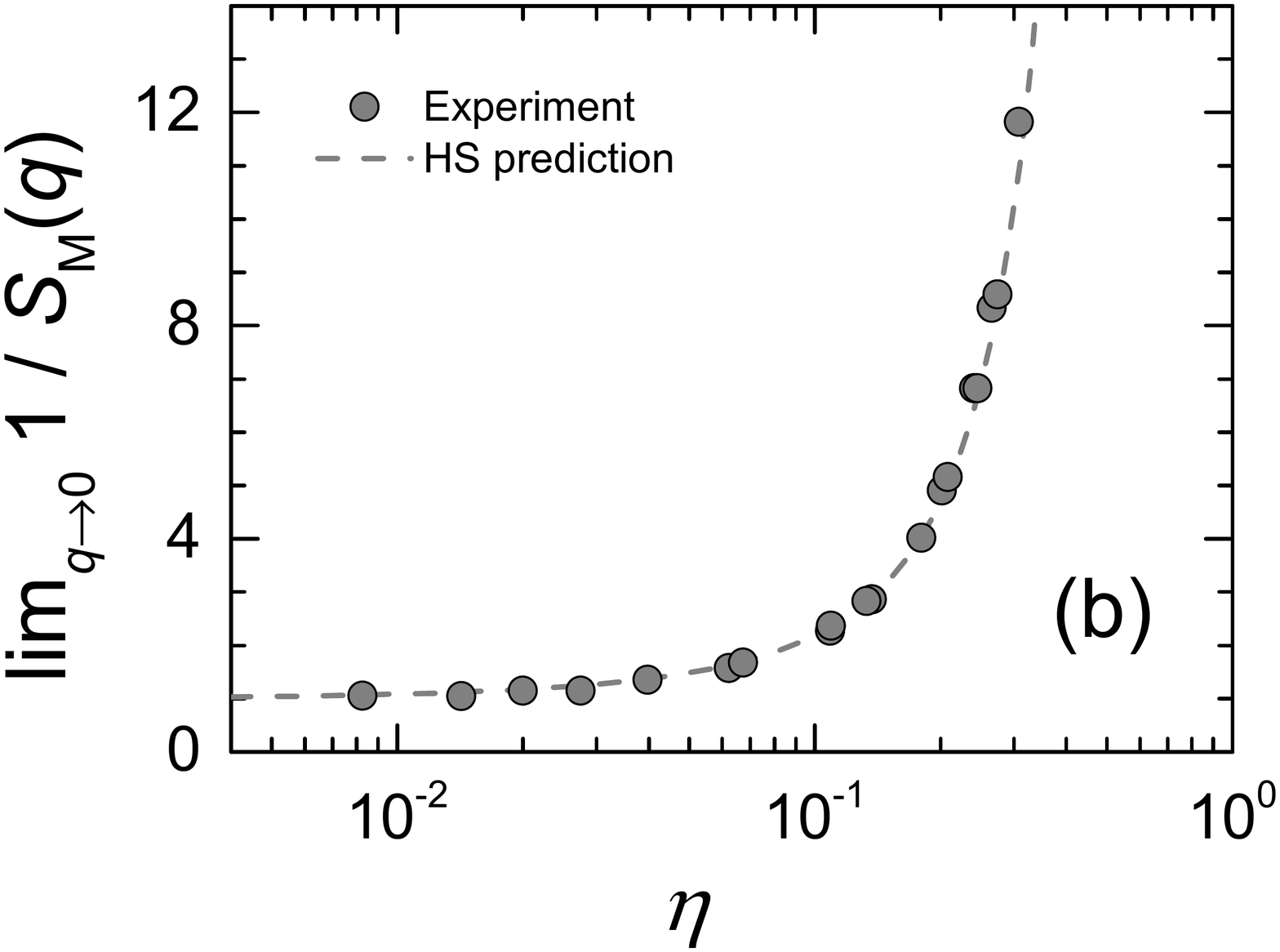}
    \end{subfigure}
	\caption{ (a) The evolution with packing fraction of the measured structure factors $\SM(q)$ for uncharged  dispersions in dodecane (particles NP1, no added AOT). The $q$-axis in the plot is logarithmic. Packing fractions are defined in the legend. The symbols denote the experimental data, while the lines are calculated polydisperse Percus-Yevick hard sphere structure factors.   (b) Comparison between the low-$q$ limit of the measured inverse structure factor (filled circles)  and the reduced isothermal compressibility $\beta / (\rhoc \chi_{\mys{T}})$ calculated from the quasi-exact Carnahan-Starling  equation of state for hard spheres (dashed line). }
	\label{fgr:SqHS}
%
%
%
%
\end{figure}

A thick polymeric shell is a highly efficient way to stabilize a nanoparticle  but  interpenetration of polymers in the shell can, particularly at high concentrations, result in a softness in the mutual interactions between grafted particles. To test if the core-shell structure of the synthesised nanoparticles was altered in concentrated dispersions we conducted a series of SAXS measurements, using the procedures outlined in Sec.~\ref{sec-SAXS}, to determine the structure factor  of uncharged particles as a function of packing fraction. The experimental results for $\SM(q)$ are depicted by the symbols shown in Fig.~\ref{fgr:SqHS}(a). The solid lines show fits to the measured $\SM(q)$ using a polydisperse hard-sphere (HS) model, using as adjustable parameters the effective HS radius $\collrad$, and the ratio $\etac/\volc$, where $\volc$ is the experimentally-determined core packing fraction. On the basis of the dilute form factor results, the size polydispersity was fixed at $\poly = 0.10$.  The HS calculations can be seen to describe the experimental structure factors extremely well over a wide range of wavevectors and nanoparticle concentrations. The hard-sphere character of the nanoparticles was confirmed further by comparing the low-$q$ limit of the measured inverse colloid-colloid structure factor $\lim_{q \rightarrow 0} 1/ \SM(q)$ with the Carnahan and Starling prediction for the isothermal compressibility of a hard-sphere fluid. The agreement evident in Fig.~\ref{fgr:SqHS}(b) is very good. Overall, we found an effective HS radius of $\collrad =$ \SI{43.3 \pm 1.3}{\nano \meter}, which is \SI{5}{\nano \meter} smaller than the magnitude of the core-shell radius $\ac + \shell = \SI{48.3}{\nano \meter}$ estimated from the  form factor analysis. We 
attribute this discrepancy to interpenetration of interlocking polymer shells at high packing fractions.  We therefore fix the effective HS radius for our systems as $\collrad = \ac+\SI{5}{\nano \meter}$. The resulting values for the effective HS radius ($\collrad$), the hydrodynamic radius determined by dynamic light scattering ($\dls$), the core radius ($\ac$) from SAXS analysis, and the normalized polydispersity for the systems studied are collected together in  Table~\ref{tblparticle}.

\begin{table}
	\caption{Nanoparticles studied\protect\footnote{$\collrad$ is the effective hard-sphere radius, $\dls$ the hydrodynamic radius, $\ac$ the core radius, and $\poly$ the normalised  polydispersity. }} \label{tblparticle}
	\begin{ruledtabular}
		\begin{tabular}{cccccc}
			Nanoparticle& $\collrad$ (\si{\nano \meter}) & $\dls$ (\si{\nano \meter}) & $\ac$ (\si{\nano \meter}) & $\poly$     \\
			\hline
			NP1 & 43.3 $\pm \; 1.3$  &  49.0 $\pm \; 2.9$  & 38.3 $\pm$ 0.3 &  0.10   \\ 
			NP2  & 58.0 $\pm \; 3.0$  & 63.0 $\pm \; 3.0$  & 53.0\footnote[2]{SAXS data not recorded, so assume $\ac = \dls - \delta$ with $\delta = $ \SI{10}{\nano \meter}.}   & 0.10   \\ 
			NP3 & 33.7 $\pm \; 1.2$  & 36.0  $\pm \; 2.5$  & 28.7 $\pm$ 0.2 & 0.13    
		\end{tabular}
	\end{ruledtabular}
\end{table}

\subsection{Charge regulation}\label{sec:mobility}

 To generate a particle charge we have used two approaches: (a) surface modification, and (b) adsorption of charged reverse micelles. Although the molecular mechanism of particle charging in low polarity solvents  is not well understood, different hypotheses have been proposed which emphasise either `charge created' on the particle by the dissociation of surface groups or `charge acquired'  by the adsorption of charged surfactant species. Our experiments use examples from both categories. System NP3 was charged by the addition of a lipophilic ionic comonomer to the  dispersion synthesis (for details, see Sec.~\ref{sec:nano}).  Dissociation of an anion from the surface of the nanoparticle generated a positive particle charge. The charge equilibrium can be represented by the single-site CR$_{1}$ dissociation process,
   \begin{equation}\label{reg:NP3}
  \ce{SN  <--> S+ + N-},
  \end{equation}
 where $S^{+}$ denotes a positively-charged surface-bound group, and $N^{-}$ a negative species.  Nanoparticle NP1 and NP2 were, in contrast, charged  negative by addition of the oil-soluble ionic surfactant Aerosol-OT at molar concentrations $\CAOT$ above the critical micellar concentration so that spherical reverse micelles form in solution. It has been proposed\cite{Roberts2008,Lee2016} that charge regulation in systems containing AOT is a multi-site CR$_{2}$ process with two independent association reactions, 
\begin{eqnarray} \label{reg:NP1}
\ce{S1 + M+ <--> S1M+}, \nonumber \\
\ce{S2 + M- <--> S2M-},
\end{eqnarray}
 where $M^{\pm}$ refer to charged reverse micelles and the balance between the two competing adsorption processes (and the net charge) depends on the hydrophilicity of the particle surface. 
 
\begin{figure}[h]
  %
  %
  %
  %
  %
  \begin{subfigure}{0.85\linewidth}
       \includegraphics[width=\textwidth]{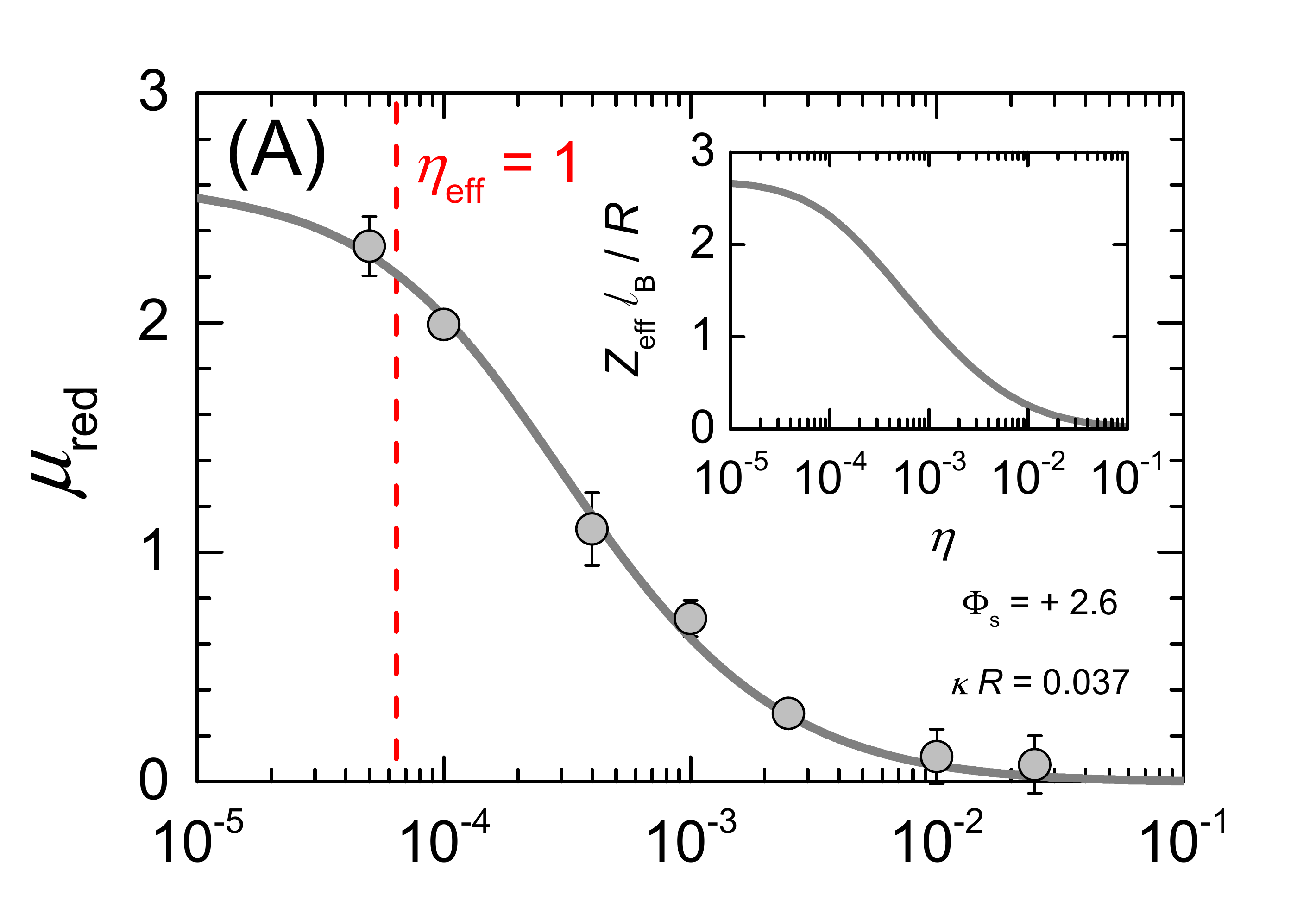}
  \end{subfigure}
  \begin{subfigure}{0.85\linewidth}
        \includegraphics[width=\textwidth]{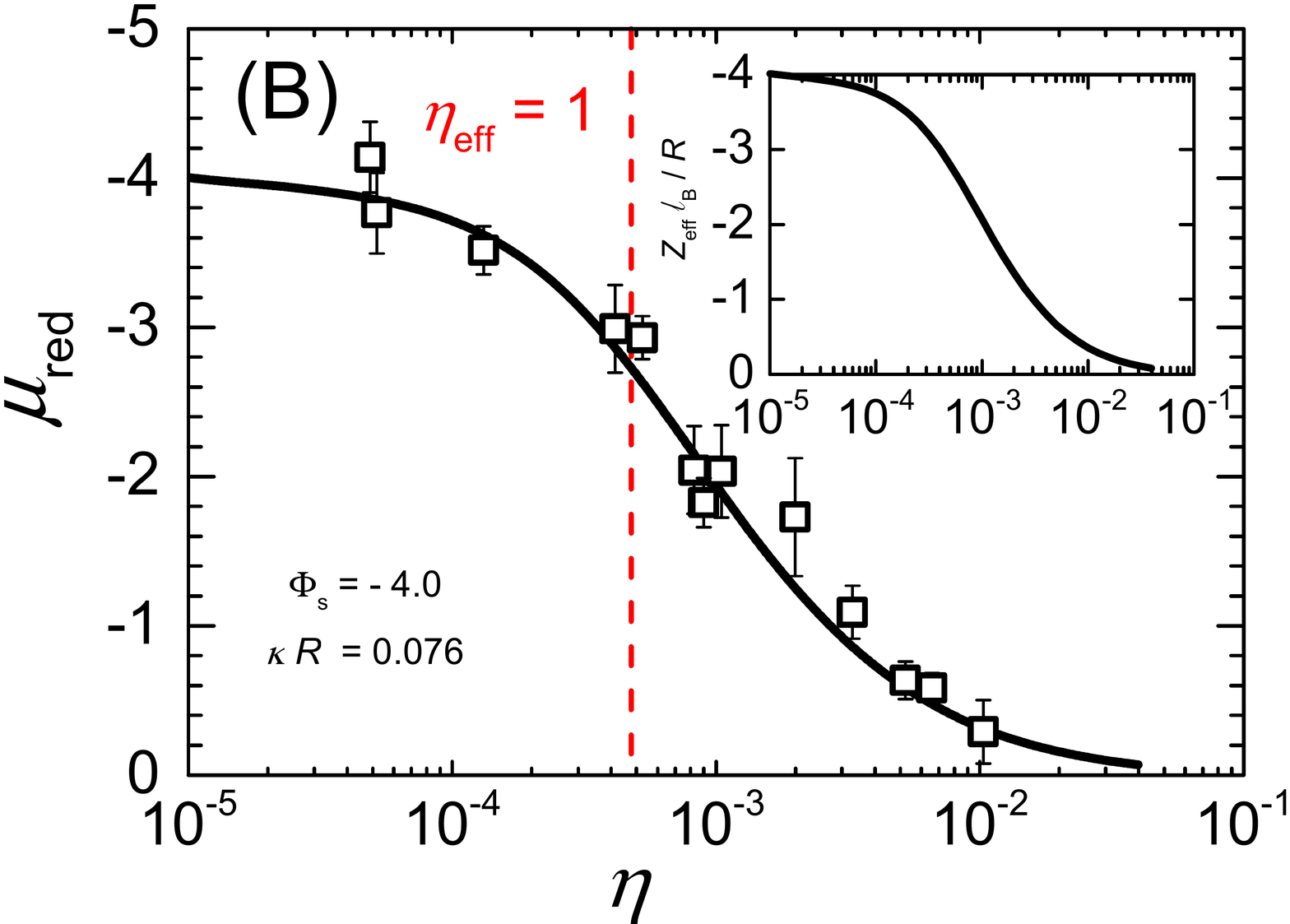}
  \end{subfigure}
  	\caption{The reduced electrophoretic mobility  $\mured$ measured as a function of volume fraction in dilute dispersions of (a) positively-charged NP3 particles (filled circles) and (b) negatively-charged NP2 particles (5 mM AOT, open squares).  The solid lines are mobilities calculated using a Kuwabara cell model (Eq~\ref{eq:mob1}) assuming a constant potential (CP) boundary condition at the surface of the particle. Fitted charge parameters are listed in Table~\ref{tblmobility}. The insets show the density dependence of the reduced particle charge predicted from the CP fit. The dashed vertical lines indicate the predictions for the concentrations where the screening clouds of neighbouring particles start to overlap.}
  	\label{fgr:density-mobility-JH}
  \end{figure}

%
%
%
  The effect of charge regulation on the surface charge was demonstrated by measurement of the reduced electrophoretic mobility $\mured$ as a function of packing fraction $\etac$, with the data shown in  Fig.~\ref{fgr:density-mobility-JH}. For a comparison between different  nanoparticle batches, we consider all mobilities in reduced units $\mured = 6 \pi \visc \lB \mu / e$ where $\visc$ is the viscosity of the solvent, and $\lB$ is the solvent-specific Bjerrum length.    For isolated particles, the reduced mobility  assumes the value $\mured^{0}= \Z \lB / \collrad = \PhiS$ in the H\"{u}ckel limit ($\kres \collrad \rightarrow 0$), where $\mured^{0}$ is the reduced mobility at infinite dilution and $\PhiS =  \beta e \phi_{\mys{s}}$ is the dimensionless surface potential\cite{Lobaskin2007}. 
  However as dispersions become more concentrated the electrophoretic mobility drops. When the electrostatic interactions are strongly screened ($\kres \collrad \gg 1$) the electrophoretic mobility shows only a relatively weak concentration dependence, since the electric-field-driven dynamics originates only from within a thin interfacial region at the particle's surface\cite{Anderson1989}. The mobility drop by a factor of $\approx 10$ as $\etac$ is increased from 10$^{-4}$ to 10$^{-2}$ seen in Figure~\ref{fgr:density-mobility-JH} is therefore quite surprising. A similar dependence of $\mured(\etac)$ was seen for all samples studied, with the data plotted in Fig.~\ref{fgr:density-mobility-JH}(a) suggesting that concentrations of $\etac \approx 10^{-5}$ are still not sufficiently low enough to reach the infinite-dilution limit $\mured^{0}$.

%
  
  Qualitatively, the sharp drop in the electrophoretic mobility $\mured$ occurs because of the strong mutual interactions between charged nanoparticles in the weak screening limit. The reduction in  $\mured$ will be significant at concentrations $\olp$ where the electrical double-layers of neighbouring particles first begin to overlap. If we approximate a charged nanoparticle and its ionic atmosphere as a new effective  particle of radius $\collrad + \kres^{-1}$ then mutual overlap will occur when the effective volume fraction $\etac_{\mys{eff}} = \etac \left[ 1 + (\kres \collrad)^{-1} \right ]^{3}$ is of order unity, or equivalently $\olp = [1+(\kres \collrad)^{-1}]^{-3}$. In the NP3 batch, where $\kres \collrad \approx 0.04$, concentration effects will be important at concentrations as low as $\etac \approx 10^{-4}$. The vertical dotted lines in Figure~\ref{fgr:density-mobility-JH} depict the concentration $\olp$ where double layer
   overlap is significant and, as evident from the plot, these lines also pretty effectively delineate the regime where $\mured$ begins to decrease. 
   
  To quantify the dramatic change in the electrophoretic mobility  we use a Kuwabara cell model, first proposed by \citet{Levine1974}, to predict $\mured$ as a function of $\etac$. We work in a spherical Wigner-Seitz cell of radius $\RWS = \collrad \etac^{-1/3}$ containing a single particle together with neutralizing co-ions and counterions. In the treatment detailed by \citet{Ohshima1997a}, which is accurate for low surface potentials and for all $\kres \collrad$ values, the reduced electrophoretic mobility $\mured$ is a function of the equilibrium electric potential $\Phi(r)$ (and its derivative) inside the cell. At  low packing fractions,  simulations have shown\cite{Everts2016,Smallenburg2011} that the particle charge predicted by either of the CR$_{1}$ and CR$_{2}$ models can be accurately mimicked by assuming a constant potential boundary condition. Fixing the packing fraction $\etac$, we solve the non-linear Poisson-Boltzmann equation varying the surface potential $\PhiS$  and the ionic strength of the reservoir  and calculate $\mured (\etac)$ (for details see Sec.~\ref{sec:vol-fract-mu}). The results of these calculations  are plotted as the solid lines in Figure~\ref{fgr:density-mobility-JH}. The calculations are seen to be in excellent agreement with the experimental data and confirm that our dispersions are indeed charge regulated. The consequences of regulation are revealed in the inset plots of  Fig.~\ref{fgr:density-mobility-JH}, where the dependence of $\Z \lB / \collrad$ on the volume fraction $\etac$ is plotted. \added[id=r18]{We found a similar level of agreement between measured and calculated electrophoretic mobilities as the screening parameter $\kappa \collrad$ was changed. The comparison between the experimental and numerical mobilities $\mured (\etac)$  as $\kappa \collrad$ was varied over almost an order of magnitude is presented in the supplementary information.}

  \begin{table*}
  	\caption{Constant-potential charging parameters: obtained from cell-model fits to experimental $\mured (\etac)$\protect\footnote{$\PhiS$ is the scaled surface potential and $\kappa \collrad$ is the effective screening parameter, obtained from a non-linear fit to Eq.~\ref{eq:mob1}}} \label{tblmobility}
  	\begin{ruledtabular}
  		\begin{tabular}{ccc|cc}
  			 \multicolumn{3}{d|}{\textit{Experimental system}} & \multicolumn{2}{d}{\textit{Fitted parameters}} \\
  			Nanoparticle & $\CAOT$ [\si{\milli \mole \per \deci \meter \cubed}] & $\kres \collrad$  & $\PhiS = e \phi_{\mys{s}} / \kBT$ & $\kappa \collrad$  \\
  			\hline
  			NP1 & 5.0 & 0.033&  $-2.0 \pm 0.1$ & 0.060 \\ 
  			NP1 & 25 &0.078&  $-2.1 \pm 0.1$ & 0.088 \\ 
  			NP1 & 50 &0.11&  $-2.4 \pm 0.1$ & 0.090 \\ 
  			NP1 & 250 &0.24&  $-2.6 \pm 0.1$ & 0.54 \\ 
  			\hline
  			NP2  & 5.0 &0.049& $-4.0 \pm 0.1$ & 0.076 \\ 
  			NP3 & 0.0  &0.0&  $+2.6 \pm 0.1$  & 0.037 
  		\end{tabular}
  	\end{ruledtabular}
  \end{table*}

\subsection{Charge at high concentrations} \label{sec:high-conc}

The low density mobility data of Fig.~\ref{fgr:density-mobility-JH} suggest that the particles should continuously discharge with increasing particle concentration. Testing this prediction in concentrated dispersions is however tricky. Light scattering measurements of $\mured$  are limited to low concentrations by multiple scattering effects.  At high $\etac$, less direct methods must be used. To this end, we have recorded the positional correlations between nanoparticles, which arise as a result of both charge and excluded volume interactions between particles, using small-angle X-ray scattering (SAXS) techniques. As we demonstrate below, careful modelling of the measured structure factor $\SM(q)$ yields a robust measure of the effective charge $\Z$.


\begin{figure}[h]
%
%
%
%
%
    \begin{subfigure}{0.49\linewidth}
    \includegraphics[width=\textwidth]{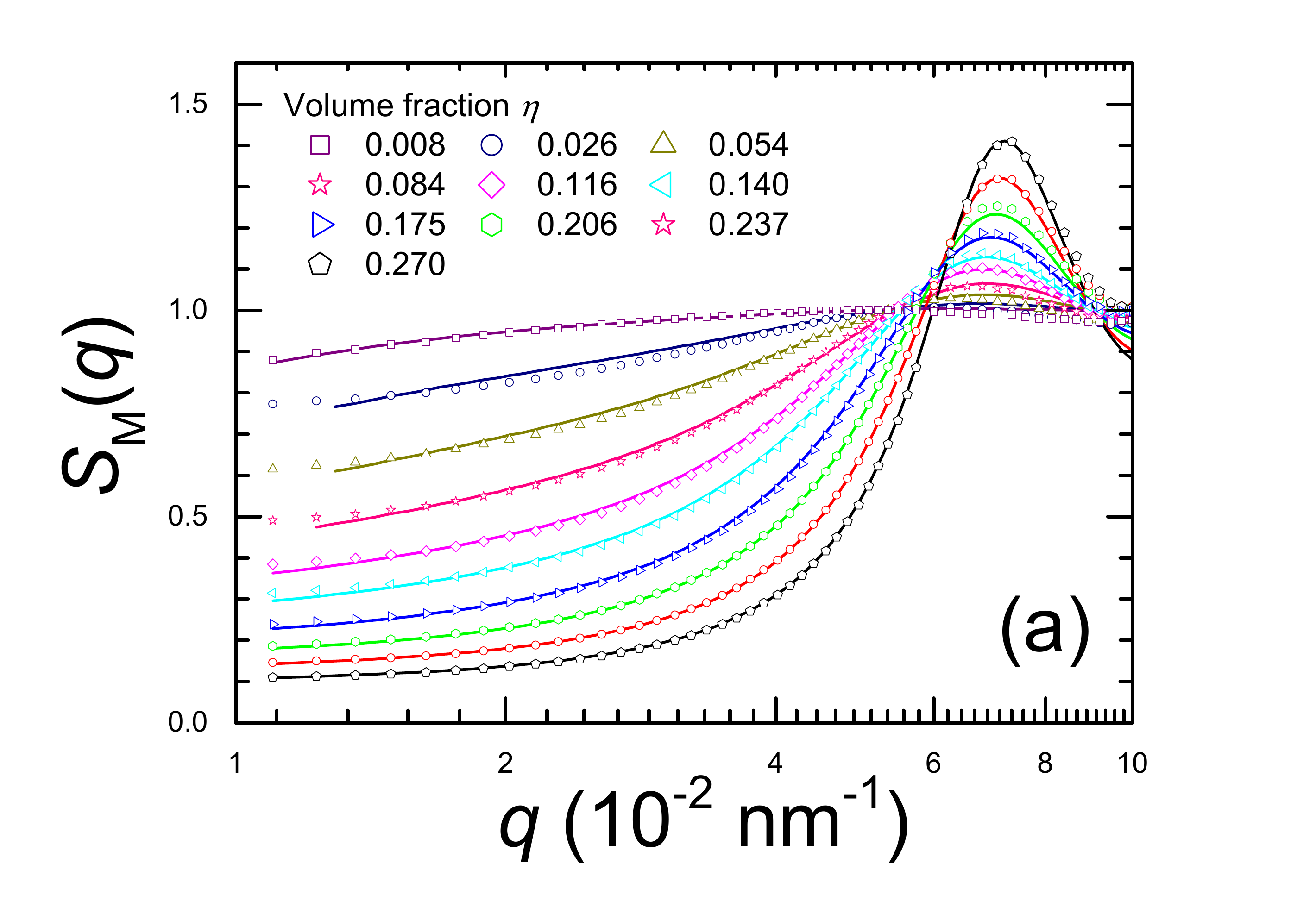}
    \end{subfigure}
    \begin{subfigure}{0.49\linewidth}
      \includegraphics[width=\textwidth]{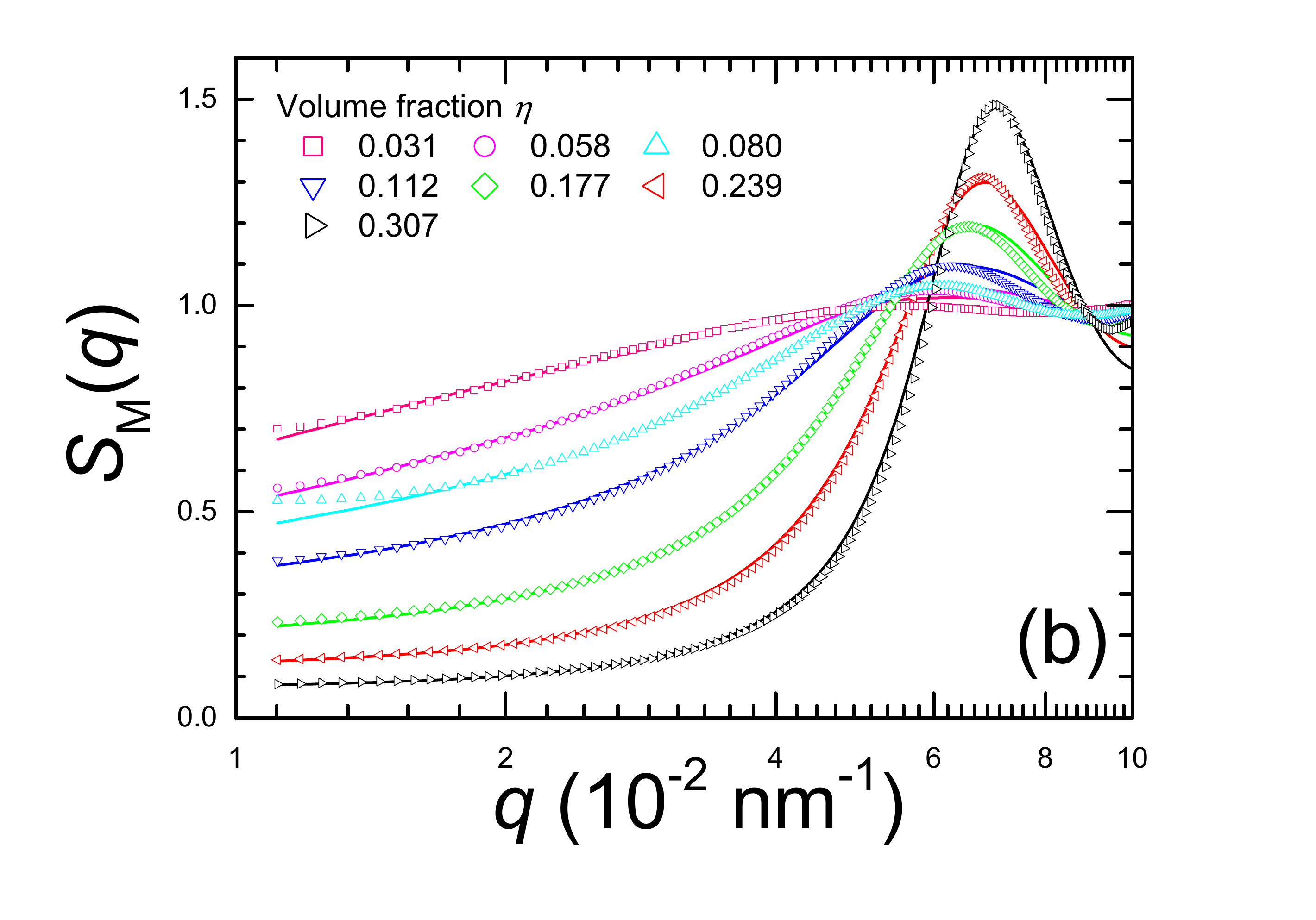}
    \end{subfigure}
    \begin{subfigure}{0.49\linewidth}
   \includegraphics[width=\textwidth]{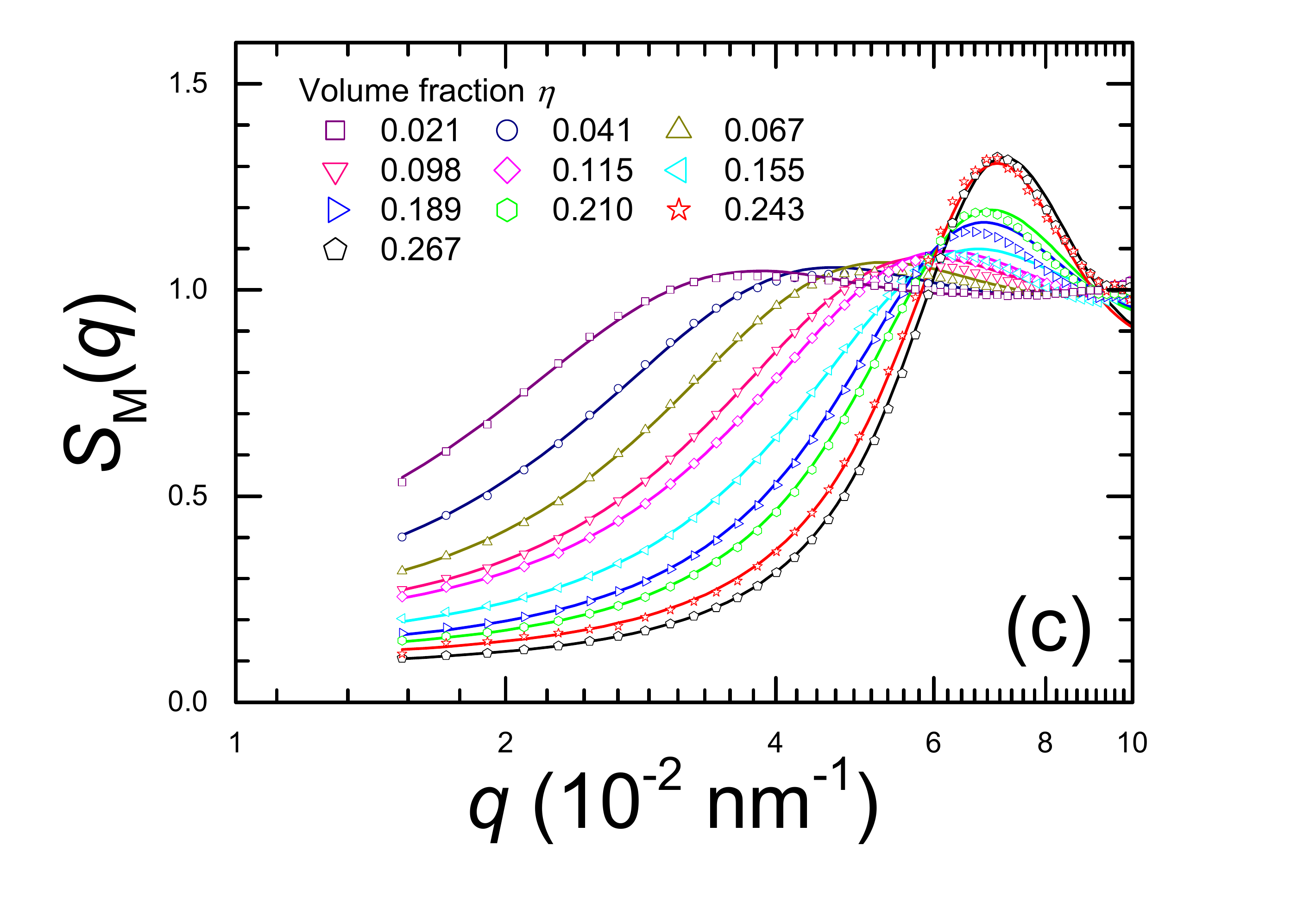}
    \end{subfigure}
    \begin{subfigure}{0.49\linewidth}
  	\includegraphics[width=\textwidth]{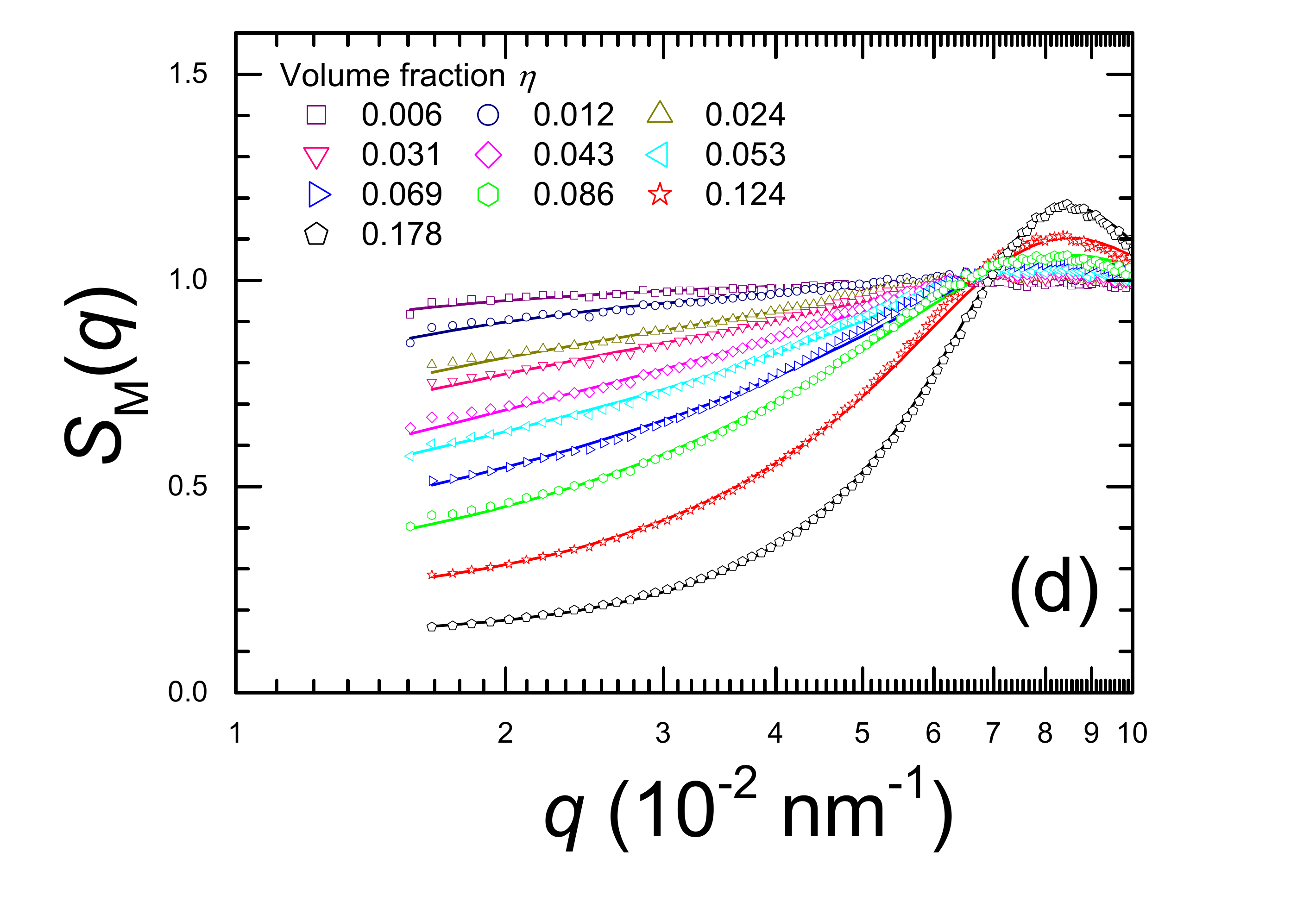}
    \end{subfigure}
    	\caption{Evolution with packing fraction ($\etac$) of measured structure factors $\SM(q)$ from four different charged PMMA systems in dodecane: (a) particles NP1 with 5 mM AOT ($\kres \collrad = 0.033$), (b) NP1 with 25 mM AOT ($\kres \collrad = 0.078$), (c) NP1 with 225 mM AOT ($\kres \collrad = 0.22$), and  (d) NP3 ($\kres \collrad = 0.0$). Packing fractions are defined in the legends. The experimental data is denoted by symbols, while the solid lines represent the best fits to $\SM(q)$, calculated from the  MPB-RMSA approximation assuming a hard-sphere Yukawa fluid. \added[id=r22]{The main peak of the structure factor at $q_{\mys{max}} \approx $ \SI{0.07}{\per \nano \meter } corresponds to particle separations of $2 \pi /q_{\mys{max}} \approx $  \SI{90}{\nano \meter }.}} \label{fgr:Sq}
\end{figure}
%

The microstructure of charge-regulating dispersions has been investigated as a function of packing fraction. The symbols in figure~\ref{fgr:Sq}(a)-(d) summarizes the measured structure factor $\SM(q)$ for four different charged systems. With increasing $\etac$, interparticle interactions start 
to dominate and the dispersions become more highly 
structured. This results in the emergence of a \replaced[id=r22]{broad nearest neighbour}{correlation}
peak at $q_{\mys{max}} \sim $ \SI{0.07}{\per \nano \meter } while at the same time, the structure factor at low $q$ decreases, behaviour that is characteristic
of a purely repulsive \added[id=r22]{fluid} system. \added[id=r22]{The lack of sharp correlation peaks in the structure factors indicates a fluid state of nanoparticles. We see no evidence for particle crystallization. Indeed since the strength of Coulombic repulsion scales linearly with size  (at the same scaled charge $\Z \lB / \collrad$, see Eq.~\ref{eq:gamma}), charge correlations are expected to be weak in a dispersion of nanoparticles with $\collrad \sim \lB$. To confirm charge interactions are not strong enough to drive crystallization at the densities studied here we refer to the phase transition data obtained recently on much larger charged colloids\cite{Kanai2015}. In the case where $\collrad \gg \lB$ the crystallization boundary was accurately modelled by the one-component plasma condition $\Gamma \ge 106$\cite{Kanai2015}, where $\Gamma$ is defined as $\Z^{2} \lB / \d$, with $\d = \rhoc^{-1/3}$ the typical spacing between particles. In scaled units, the coupling constant is $\Gamma =  \left( \frac{3 \etac}{4 \pi}\right)^{1/3} \left( \Z \lB /\collrad \right)^{2}  \cdot \left( \collrad / \lB \right) $. Using values of $|\Z \lB /\collrad | \approx 2$ and $\collrad / \lB \approx 2$ we estimate $\Gamma \approx  3$ for our nanoparticle system at $\etac \approx 0.3$, confirming the conclusion from the $\SM(q)$ data that all of the suspensions studied here are disordered fluids. }

\begin{figure}[h]
	%
	%
	\centering
		\includegraphics[width=0.45\textwidth]{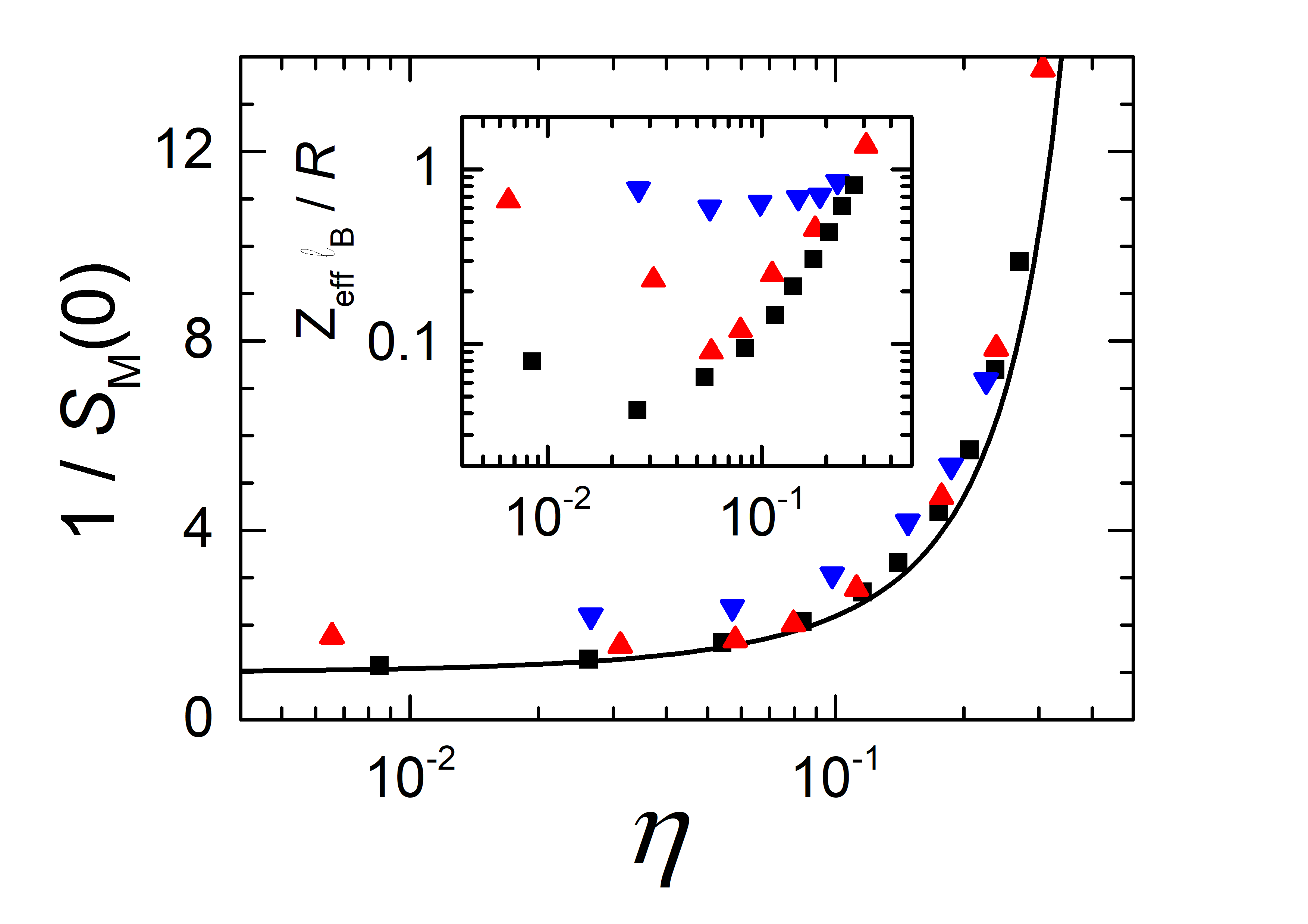}
	\caption{Low-$q$ limit of the inverse structure factor,  $\lim_{q \rightarrow 0} 1/ \SM(q)$, for charged dispersions as a function of colloid packing fraction $\etac$. Squares (black) 5mM AOT, up triangles (red) 25 mM AOT, down triangles (blue) 50 mM AOT. Solid line indicate reduced isothermal compressibility $\beta / (\rhoc \chi_{\mys{T}})$ calculated from Carnahan-Starling  equation of state for hard spheres. Inset scaled effective charges $\Z \lB / \collrad$ calculated from experimental low-$q$ data using MSA (see text for details).}
	\label{fgr:sq-compress}
\end{figure}

To begin our analysis of the scattering data, we determine the concentration dependence of the inverse isothermal osmotic compressibility   $\chi_{\mys{T}}^{-1} = \rhoc \left(  \frac{\partial\Pi}{\partial\rhoc}\right)_{T,\mys{salt}}$. Here $\Pi$ is the osmotic pressure of the suspension measured at a colloid number density of $\rhoc$ and the derivative is evaluated at a constant chemical potential of salt\cite{Dobnikar2006}. From the Kirkwood-Buff relation\cite{Kirkwood1951}  the infinite-wavelength limit of the colloid-colloid structure factor,  $S(0) = \lim_{q \rightarrow 0}S(q)$,  is related to $\chi_{\mys{T}}$ by the identity,
\begin{equation}\label{eq:KB}
\chi_{\mys{T}}^{-1} = \frac{\rhoc \kBT}{S(0)},
\end{equation}
which is exact for a monodisperse suspension. Experimentally, we determined the low-$q$ limit by extrapolating a  linear plot of $\SM(q)$ versus $q^{2}$ to $q=0$. The resulting  estimates of the 
reduced inverse osmotic compressibility $\beta / (\rhoc \chi_{\mys{T}}) = 1/\SM(0)$  as a function of the packing fraction $\etac$ for dispersion NP1 with AOT concentrations of $\CAOT = $ 5, 25 and \SI{50}{\milli \mole \per \deci \meter \cubed } are plotted in Figure~\ref{fgr:sq-compress}. At each volume fraction and AOT concentration, the effective particle charge $\Z \lB / \collrad$ was determined using a mean spherical approximation (MSA) expression for the structure factor of a HSY fluid, recently derived by \citet{Vazquez-Rodriguez2016}. The only free parameter in the computation is $\Z$, since the screening length $\kap^{-1}$ is fixed by $\Z$ (Eq.~\ref{NP1,NP2})  and the self-ionization equilibrium constant $K$ is known from previous work\cite{Roberts2008}. 

The inset plot in Figure~\ref{fgr:sq-compress} shows the particle charges computed from the long-wavelength limit of the structure factor. Surprisingly, we see substantial disagreements at intermediate packing fractions from the CP predictions, as is evident from a quick comparison between Figures~\ref{fgr:extrapolated-point-charges} and \ref{fgr:sq-compress}. The effective charge does not decrease smoothly with increasing $\etac$ but  instead displays a minimum at $\etac \approx 0.06$ before finally \textit{increasing} in magnitude with particle density.

To confirm the validity of the trends identified above, we have  determined the particle charge $\Z(\etac)$ using a second independent technique. The full-$q$ dependence of the measured structure factor $\SM(q)$ was compared to a $S(q)$ calculated using the quasi-exact MPB-RMSA integral equation scheme\cite{Heinen2011}, assuming a HSY effective pair potential. 
%
The effect of size polydispersity was neglected since polydispersity indices  are small (see Table~\ref{tblparticle}) and $\SM(q)$ contains no sharp peaks. In all, two adjustable parameters were used in our theoretical modelling of $\SM(q)$: the scaled particle charge $\Z \lB / \collrad$ and the ratio $\etac / \volc$, where $\volc$ is the experimentally assigned core volume fraction.  The screening parameter $\kap \collrad$ was derived self-consistently from the fitted values for $\Z \lB / \collrad$ and $\etac$ using Eq.~\ref{NP1,NP2}. To check the reliability of this analysis, we first fitted the structure factors measured in the uncharged system (without any added AOT). 
The fitted charge was, as expected, close to zero with an average of $\left < \Z \lB / \collrad \right > = 0.05 \pm 0.05$, validating our approach.

\begin{figure}[h]
	
	%
	%
	%
	%
	%
	%
	\centering
	\includegraphics[width=0.45\textwidth]{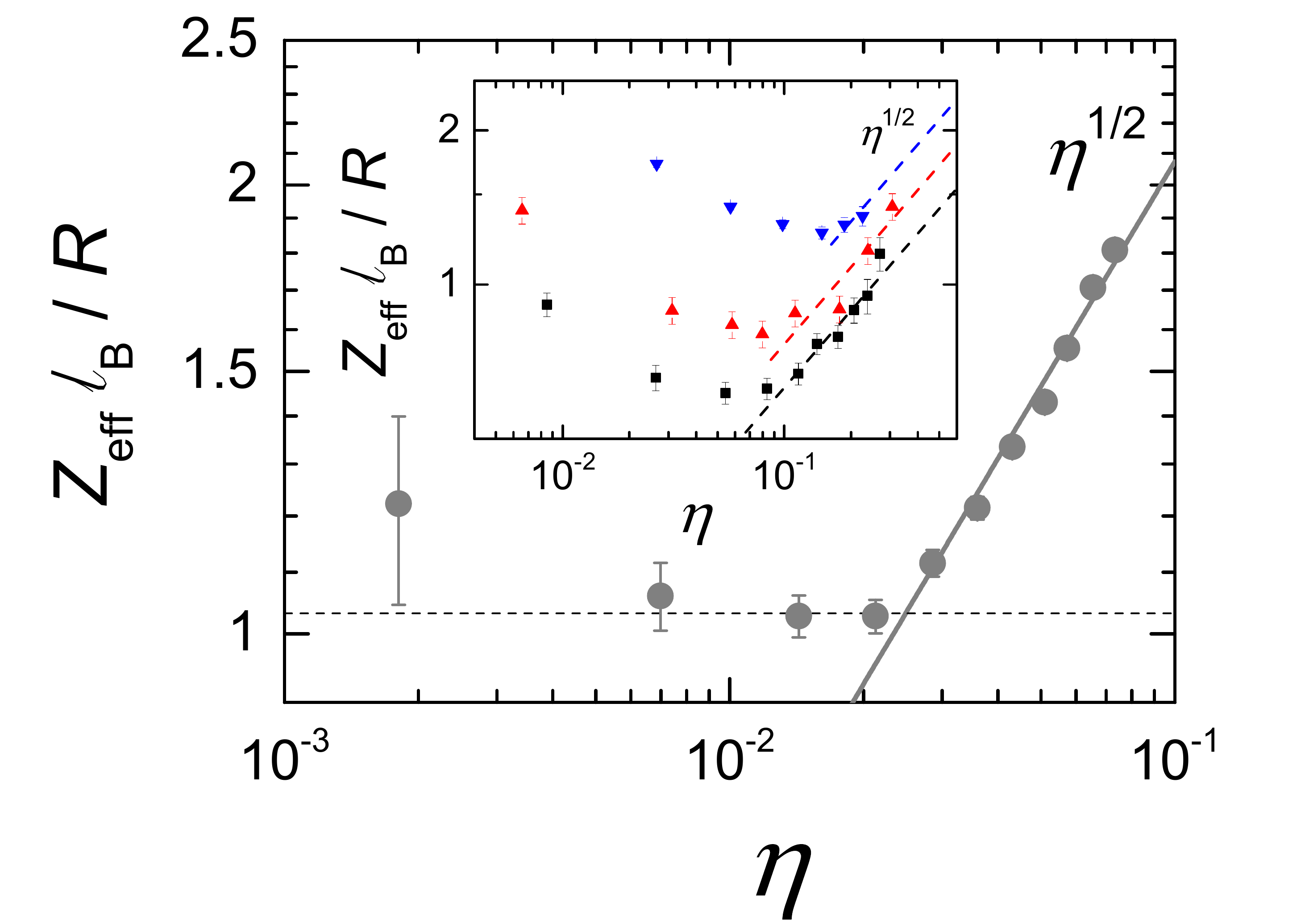}
	\caption{Measured charge versus volume fraction, for different salt conditions. The scale in the plot is logarithmic. Main figure contains data for no-salt system (NP3, circles). Inset show experimental data for NP1 dispersion, with  salt concentration increasing from bottom to top.  Symbols same as Fig.~\ref{fgr:sq-compress}.}
	\label{fgr:charge_phi}
\end{figure}

Reassured by the accuracy of our fitting strategy, $\Z$ was determined for the charged systems. The resulting particle charges as a function of $\etac$, are reproduced in Fig.~\ref{fgr:charge_phi}. The increase in the effective charge at high densities, evident in the compressibility data, is particularly clear for the counterion-only system (NP3, main body of Fig.~\ref{fgr:charge_phi}). The inset plot confirms that the same qualitative trend persists when small amounts of salt are present (system NP1). Remarkably, we find that all of the experimental data is consistent with an asymptotic scaling of the particle charge, $\Z(\etac) \sim \sqrt{\etac}$, at high packing fractions. The evidence for this expression is strongest from the data collected on the no-salt system (NP3) although a $\sqrt{\etac}$ dependence is also clearly visible in the NP1 dataset, particularly at low $\CAOT$. Finally, we note that the onset of the $\sqrt{\etac}$ scaling regime  moves to higher packing fractions with increasing background salt concentrations. So, for instance, in the no-salt system the charge starts to increase at $\etac \approx 0.02$, while at $\kres \collrad = 0.033$ the scaling regime is delayed until  $\etac \approx 0.07$, $\etac \approx 0.10$ for $\kres \collrad = 0.078$, and $\etac \approx 0.20$ for $\kres \collrad = 0.11$. Consistent with this picture, at the highest salt concentration ($\CAOT = \SI{225}{\milli \mole \per \deci \meter \cubed }$) where $\kres \collrad =0.22$, the charge decreases monotonically with increasing $\etac$ (data not shown) and there is no sign of any charge minimum.

\subsection{Origins of charge increase}\label{sec:charge-origin}

\begin{figure}[h]
%
%
%
 	\centering
 		\includegraphics[width=0.45\textwidth]{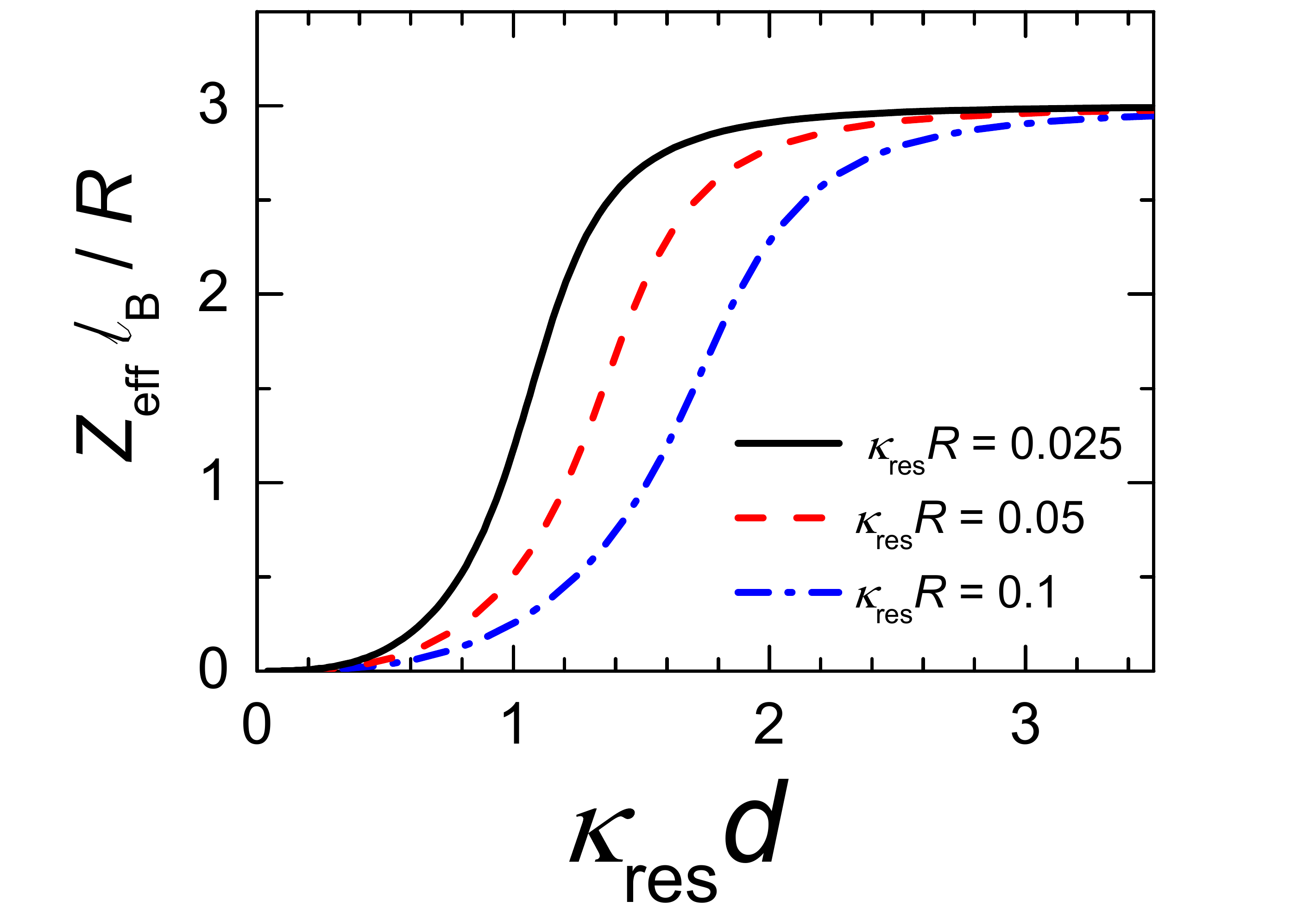}
 	\caption{Charges predicted by constant potential model as a function of the particle-to-particle separation $d$, in units of the Debye length $\kres^{-1}$, for several screening parameters $\kres \collrad$.   The surface potential $\PhiS$ is chosen so that the charge in the widely-separated limit is $\Z \lB / \collrad =3$. }
    \label{fgr:charge-calc}
\end{figure}

\added[id=r22]{We emphasise that the pronounced minimum in the effective charge $\Z(\etac)$ seen in Fig.~\ref{fgr:charge_phi} is not caused by a change in the microstructural order of the suspension, such as crystallization. First, the height of the main peak in the structure factor $\SM(q_{\mys{max}})$ is significantly smaller than the Hansen-Verlet criterion of $\SM(q_{\mys{max}})=2.85$ for freezing\cite{Hansen1969} even after allowing for polydispersity\cite{Kesavamoorthy1989}, second the structure factor contains no sharp peaks indicative of crystal formation, and finally comparison with the phase boundaries measured in suspensions of much larger-sized charged particles\cite{Kanai2015} suggest our samples lie deep within the fluid phase of charged particles.}  To \replaced[id=r22]{rationalize }{ understand the reasons for } the increase in  \replaced[id=r22]{$\Z$}{ the particle charge seen} at high densities we go back and reconsider the consequences of constant potential boundary conditions, recognizing that the CP model fairly accurately mimics charge regulation. The reduced particle charge $\bare \lB / \collrad$ is a dimensionless quantity, so it can depend  only on combinations of dimensionless parameters. In the usual description\cite{Smallenburg2011}  of a dispersion of constant-potential spheres the set of dimensionless variables is chosen as the packing fraction $\etac$, the screening parameter $\kres \collrad$, and the scaled surface potential $\PhiS \equiv e \phi_{\mys{s}} /  \kBT$. However, a more intuitive picture emerges if we use the average separation $\d$ between particles in units of the Debye length $\kres^{-1}$,  in place of $\etac$. \deleted[id=r22]{We choose the mean center-to-center separation $\d$ between particles as $\d = \rhoc^{-1/3}$ or equivalently,
$\d = \left( \frac{3 \etac}{4 \pi} \right)^{-1/3} \collrad$}
 The dimensionless  ratio  $\lambda \equiv \kres \d$  details the extent of the overlap between the electrical double layers surrounding neighbouring particles, with large $\lambda$ corresponding to widely separated particles. $\lambda$ dictates the curvature of the electrostatic potential between particles. From Gauss's law, the charge on the surface of the nanoparticle is 
\begin{equation}\label{eq:gauss}
\frac{\bare \lB}{\collrad} = - \collrad \Phi^{'}(\collrad),
\end{equation}
where the prime denotes a derivative with respect to the radial variable. So a smaller value of $\lambda$ will correlate naturally with a smaller particle charge. Numerical solutions of the CP model for $\bare \lB / \collrad$ as a function of $\lambda$, plotted in Figure~\ref{fgr:charge-calc},    confirm this picture. When particles are widely spaced the colloidal charge plateaus at an asymptotic value, which in linear-screening theory is 
\begin{equation}\label{eq:asmptotic}
\frac{\bare \lB}{\collrad} = \PhiS (1+\kres \collrad).
\end{equation}
Moving the particles closer together results in progressive overlap of the counterion atmospheres around each particle and subsequent colloid discharge, with $\bare \lB / \collrad \rightarrow 0$ as $\lambda \rightarrow 0$. Furthermore, Figure~\ref{fgr:charge-calc} reveals that the charge dependence $\bare(\lambda)$  is only weakly dependent on $\kres \collrad$, under the weak-screening conditions ($\kres \collrad \ll 1$) appropriate to our experiments. 

%
%

Before we compare directly the experimental results and cell model predictions we emphasise a key distinction between the two approaches. The cell model calculations, in common with most theory and simulation studies, utilize a \textit{grand-canonical} treatment of the electrolyte in which the charged dispersion is assumed to be osmotically-coupled to an external ion reservoir with a constant chemical potential, so that calculations are performed at a fixed value of $\kres \collrad$. By contrast, in experiments there is no ion reservoir. Measurements are conducted in a \textit{canonical} ensemble, in which the equilibrium concentration of mobile ions and hence $\kap \collrad$ varies with both the packing fraction $\etac$ and charge $\Z$. To estimate the density-dependent screening in experiments we assume all ions are univalent and use Eq.~\ref{eq:Denton}, or its equivalent in scaled units, 
\begin{equation}\label{eqkappa}
(\kap \collrad)^{2} = (\kres \collrad)^{2} + 3  \etac \left | \frac{\Z \lB}{\collrad} \right |.
\end{equation}
The dimensionless range $\leff = \kap d$  is accordingly,
\begin{equation}\label{eq:lambda}
\leff = \left(\frac{3 \etac} {4 \pi}\right)^{-1/3} \left[(\kres \collrad)^{2} + 3 \etac   \left|\frac{\Z \lB}{\collrad} \right|                   \right]^{1/2}.
\end{equation}
Inspection of Eq.~\ref{eq:lambda} reveals a subtle dependence of $\leff$ on particle concentration. In the conventional salt-dominated regime, where the added electrolyte exceeds the number of counterions released from the surface of the particles, the overlap of the double layers between particles \textit{grows} ($\leff \sim \etac^{-1/3}$ reduces) as the dispersion is concentrated. This behaviour is however reversed in the no-salt limit, where the right-hand term of Eq.~\ref{eq:lambda} dominates. In this  regime,  $\leff \sim \etac^{1/6}$ so that, rather counter-intuitively,  colloids become less strongly interacting at high packing fractions. The cross-over from a salt-dominated to a counterion-dominated screening regime  occurs at a particle volume fraction $\cross$ which may be identified with the location of the turning point, where $\partial \leff / \partial \etac = 0$. Differentiation of Eq.~\ref{eq:lambda} yields the estimate for the  cross-over packing fraction $\cross$,
\begin{equation}\label{eq:crossover}
\cross = \frac{ 2(\kres \collrad)^{2}}{ 3\left | \Z \lB / \collrad \right|},
\end{equation}
above which we enter the counterion-dominated regime. Since typically $\left | \Z \lB / \collrad \right | \approx \mathcal{O} (1)$ Eq.~\ref{eq:crossover} reveals that the counterion-dominated regime will be accessible only if $\kres \collrad \ll 1$.

\begin{figure}[h]
	
	%
	%
	%
	%
	%
	%
        \centering
      		\includegraphics[width=0.45\textwidth]{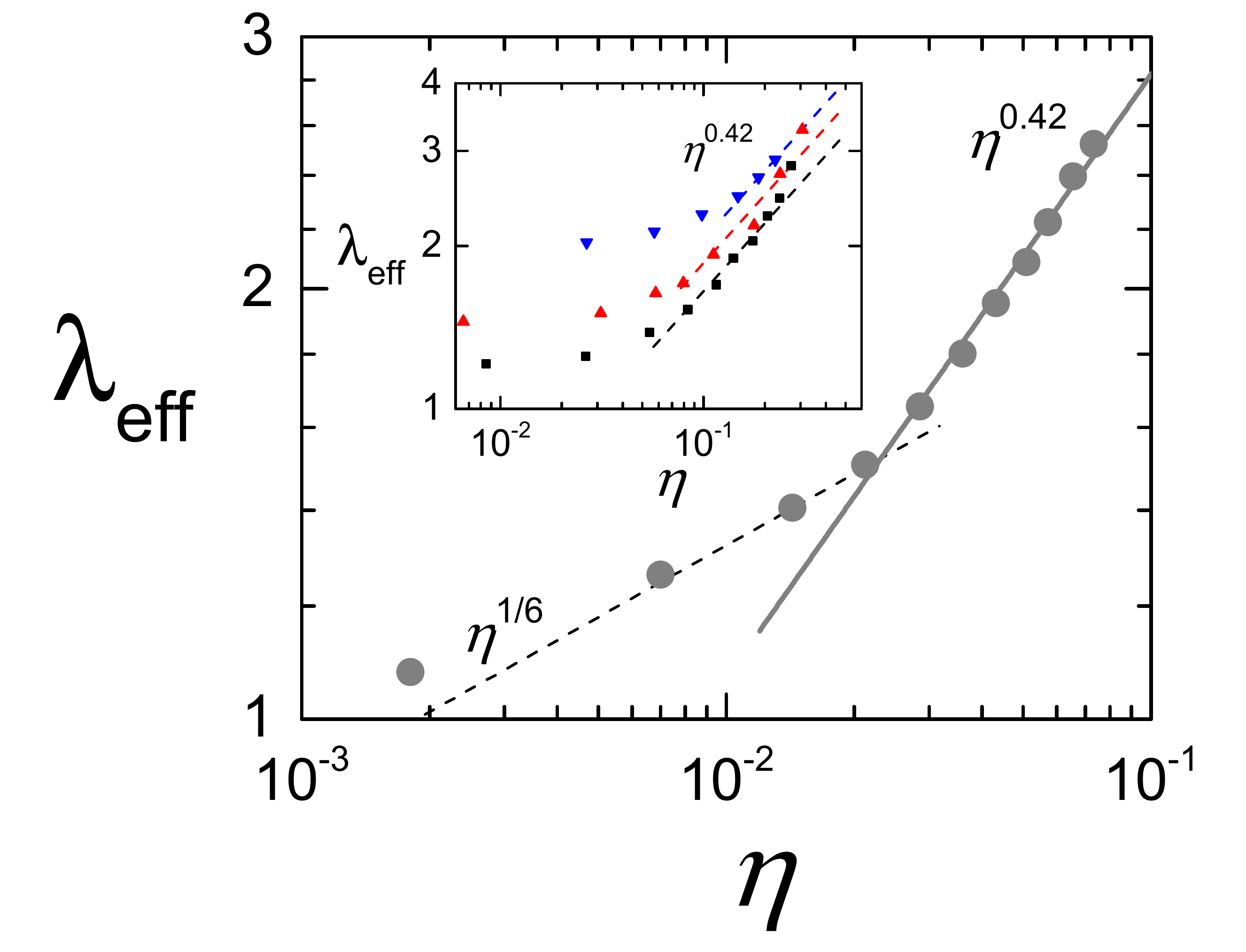}     
    	\caption{Measured  particle-to-particle separation (circles) in no-salt (NP3) dispersions,  as a function of packing fraction $\etac$. Inset shows NP1 data in presence of added salt. Symbols same as Fig.~\ref{fgr:sq-compress}.}
    	\label{fgr:centre-centre-lambda}
\end{figure}

To quantify in which screening regime our experiments lie, we determine the dimensionless range  $\leff$  as a function of $\etac$ using the particle charge $\Z$ measured by SAXS (Fig.~\ref{fgr:charge_phi}), \added[id=r22]{and } the in-situ screening parameter $\kap \collrad$ calculated from Section~\ref{sec:mat-kappa}\deleted[id=r22]{, and the interparticle spacing $d$ calculated from Eq.~\ref{eq;spacing}}. Figure~\ref{fgr:centre-centre-lambda} shows the resulting packing fraction dependence for the no-salt system (NP3, circles),  and for the NP1 system  at AOT concentrations of $\CAOT = $ 5 (squares), 25 (up-triangles) and \SI{50}{\milli \mole \per \deci \meter \cubed } (down-triangles) in the inset plot.  We identify three interesting features. First, Fig.~\ref{fgr:centre-centre-lambda} reveals that $\leff$ increases with increasing $\etac$, so our SAXS measurements all lie within the counterion-dominated screening regime. This conclusion agrees with the cross-over volume fractions estimated from  Eq.~\ref{eq:crossover}.  So, for instance, the minimum charge measured in the $\CAOT = \SI{50}{\milli \mole \per \deci \meter \cubed }$ ($\kres \collrad \approx 0.11$)  system  is $\left | \Z \lB / \collrad \right| \approx 1.3$ and Eq.~\ref{eq:crossover} predicts $\cross \approx 0.006$, while SAXS experiments were performed at concentrations of $\etac > 0.02$.  Second, the low-$\etac$  data is consistent with the $\lambda \sim \etac^{1/6}$  dependence expected for the counterion-only limit. This is particularly evident in the experimental data for the NP3 system, where the background ion concentration is  close to zero. Finally, we note that this asymptotic dependence only holds for low $\etac$: at intermediate and high volume fractions a new power-law regime with $\lambda \sim \etac^{0.42 \pm 0.01}$ appears. This is most clearly seen in the counterion-only data.  We note from Eq.~\ref{eq:crossover} that a power-law of $\lambda \sim \etac^{0.42 \pm 0.01}$ suggests that the particle charge must scale as $\Z \sim \etac^{0.51 \pm 0.02}$ for a no-salt system, consistent with Fig.~\ref{fgr:charge_phi}.

\begin{figure}[h]
	%
	%
	%
	\centering
		\includegraphics[width=0.45\textwidth]{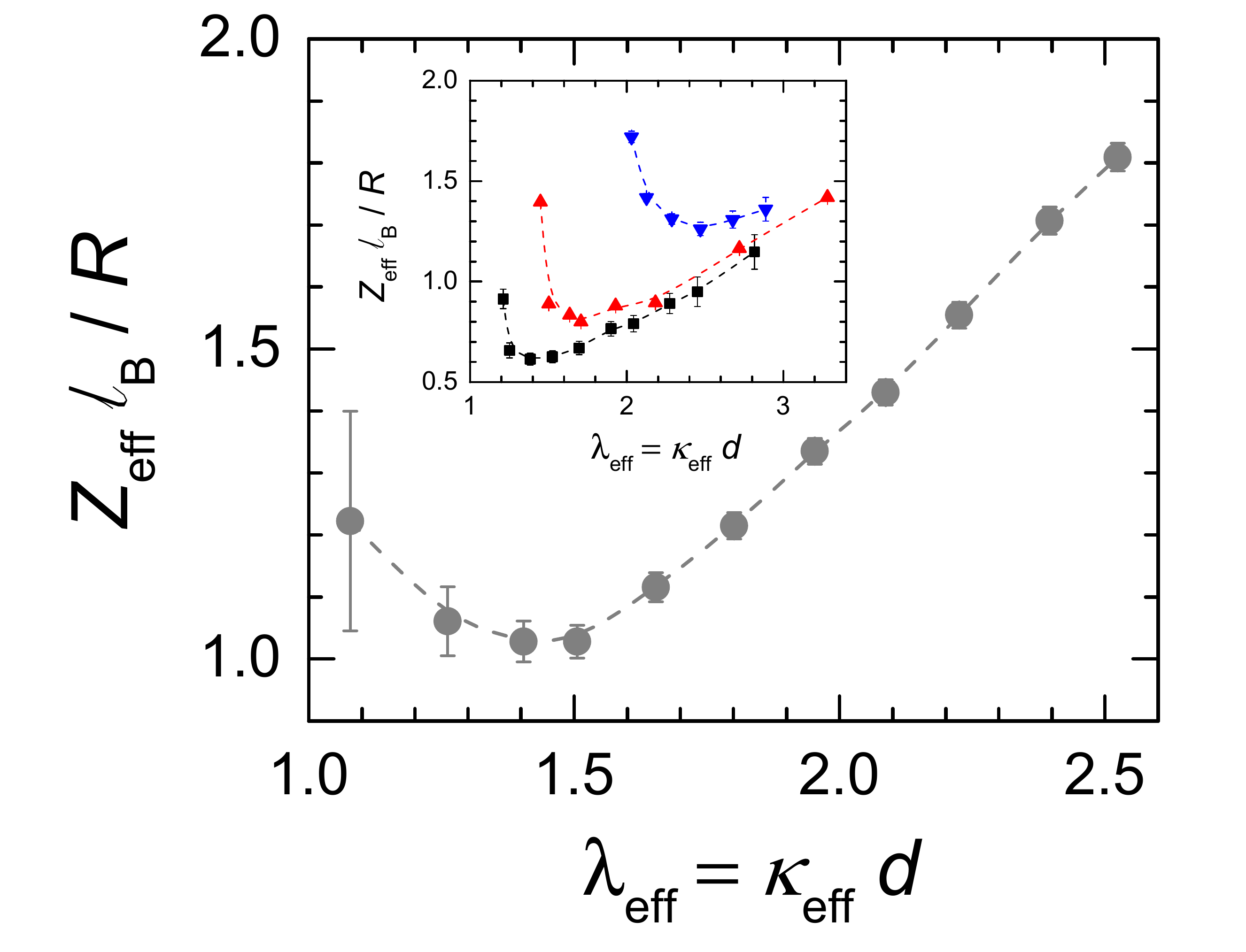}
	\caption{Experimental charges replotted as a function of the particle-to-particle separation $d$, in units of the Debye length $\kres^{-1}$. Main figure contains data for counterion-only system (NP3, circles). Inset show experimental data for NP1 dispersion, with  salt concentration increasing from bottom to top.  Symbols same as Fig.~\ref{fgr:Sq}. }
	\label{fgr:charge-lambda}
\end{figure}

The observation that the SAXS data lie within the counterion-dominated screening regime rationalizes qualitatively why the measured particle charge increases with increasing particle density.  Numerical solutions of the nonlinear Poisson-Boltzmann equation show that CP-particles should charge-up (see Fig.~\ref{fgr:charge-calc}) as the range of the electrostatic interactions becomes significantly smaller than a typical particle spacing. In the no-salt limit, the thickness $1/\kap$ of the double layer shrinks faster with increasing colloid  concentration ($1/\kap \sim \etac^{-1/2}$) than the mean spacing $\d$ between particles  ($\d \sim \etac^{-1/3}$) so that as the packing fraction is increased, charged particles become less highly correlated, $\leff$ grows, and the particle charge should accordingly be boosted. To facilitate a direct comparison, we plot in Fig.~\ref{fgr:charge-lambda} experimental values for the dimensionless particle charge $\Z \lB / \collrad$ as a function of the range $\leff$ of the interactions. The data are represented by the symbols and should be compared to the CP-model predictions plotted in Fig.~\ref{fgr:charge-calc}. Fig.~\ref{fgr:charge-lambda} confirms that $\Z(\leff)$ is an increasing function of $\leff$, at least at large $\leff$, in broad agreement with CP predictions. However,  a pronounced minimum appears in the experimental data at $\leff \approx 1.5$, which is not predicted by the cell model. This may be a consequence of the neglect of three-body and higher order correlations\cite{Klein2002,Russ2002} in a spherical approximation. As $\leff$ is reduced, each charged particle interacts not only with its immediate shell of neighbours but increasingly with the next nearest neighbour shell. Three-body interactions between charged particles are attractive within Poisson-Boltzmann theory\cite{Russ2002}, a fact which has been interpreted in terms of electrostatic screening by macro-ions\cite{Klein2002}. We speculate that this many-body mechanism of charge screening, which will be strongest at small $\leff$,  could enhance the dissociation of surface charge groups and so increase the particle charge above the predictions of a simple spherical approximation.

%

\section{Conclusions} \label{sec:conclude}

We have determined the density dependence of the structure factor $\SM(q)$ in weakly-charged nonpolar colloidal dispersions  using small-angle scattering techniques. An extended range of particle number densities was employed, with colloid packing fractions varying by a factor of approximately $10^{3}$.  By utilizing small radii nanoparticles and low ionic strengths, typical of nonpolar systems, we ensure  measurements remain in the weak screening regime ($\kres \collrad \ll 1$) even up to packing fractions approaching 30\%. The effective charge $e\Z$ of the particles was determined from a comparison between $\SM(q)$ and structure factors calculated from the highly-accurate modified penetrating-background corrected rescaled mean spherical approximation (MPB-RMSA) introduced by \citet{Heinen2011}. The Debye screening parameter $\kap$ was calculated self-consistently from the effective charges and the measured background ion concentrations. The resulting $\Z (\etac)$ values are in near quantitative agreement  with effective charges measured by electrophoresis, at least for low packing fractions where mobility measurements are feasible. Our results for $\Z(\etac)$ cover a broader range of packing fractions than  previous studies and allow us to scrutinize in great detail predictions for the density-dependence of electrostatic interactions in concentrated charged dispersions.

This work suggests that the Derjaguin-Landau-Verwey-Overbeek (DLVO) theory widely used to describe the interaction between two uniformly-charged particles, cannot describe nonpolar colloids. We test this conclusion by examining two different experimental systems, in which charge was generated by, either the dissociation of surface-bound groups or, alternatively by the adsorption of charged surfactant micelles. DLVO theory assumes a fixed spherically-uniform charge distribution. The results presented here clearly demonstrate that such an assumption does not hold in the case of nonpolar dispersions where colloids are charge-regulated and surface charge density varies as a function of packing fraction. We find that a charge regulating system is less repulsive than an equivalent dispersion with a fixed charge distribution. This is due to a reduction in the dissociation of weakly-ionizing groups as their separation decrease, driven by the need to avoid a local increase in the counterion density in the region between neighbouring colloids. 

Theoretical models of charge regulation\cite{Everts2016,Smallenburg2011} predict a strong decrease in the effective charge upon increasing particle concentration. While we find that $\Z$ does indeed decrease at low packing fractions, our data reveals unexpectedly a pronounced minimum  in the effective charge at $\etac \approx \cross$. For $\etac > \cross$,  the effective charge  \textit{increases} with particle concentration in apparent disagreement to existing models.  We observe in all samples a near square-root scaling of the measured effective charge, $\Z \sim \etac^{1/2}$, at high densities. The packing fraction at the charge minimum is $\cross \approx 10^{-2}$ in the case of a no-salt system and  shifts to progressively higher volume fractions (reaching $\cross \approx 10^{-1}$ at $\kres \collrad = 0.22$) as the salt concentration in the system is increased. 

\begin{figure}[h]
	\centering
	\adjustbox{trim={0.25\width} {0.07\height} {0.3\width} {0.0\height}, clip,width=0.5\textwidth} %
        {\includegraphics[]{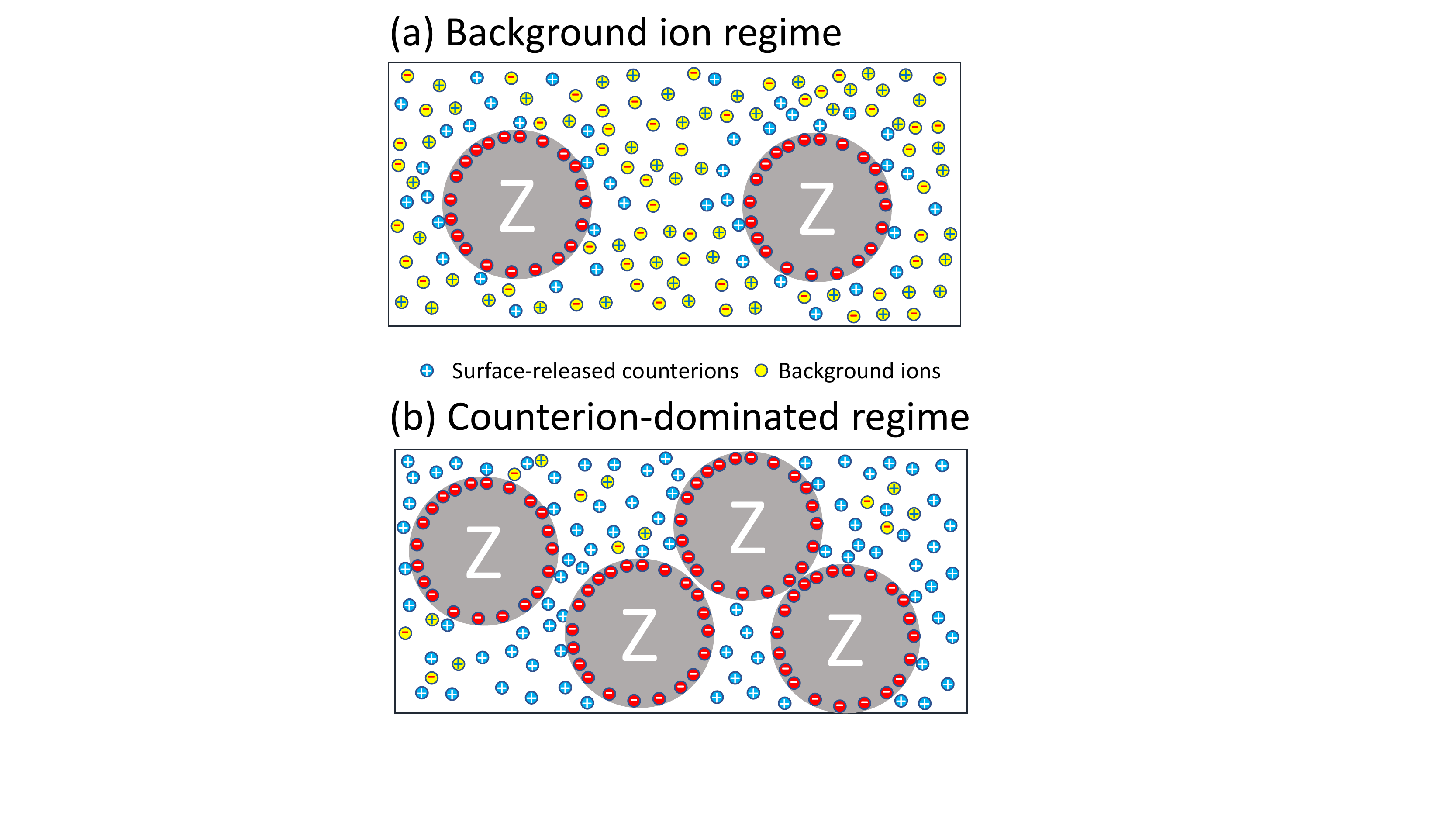}}
	\caption{Illustration of background-ion and counterion-dominated screening regimes. (a) At low nanoparticle packing fractions, the majority of the ions in solution are contributed by the bulk electrolyte (shown in yellow). (b) At high $\etac$, most of the ions in solution arise from the dissociation of surface groups on the nanoparticles (shown in blue). }
    \label{fgr:cartoon}
\end{figure}

Our findings can be explained \deleted[id=r23]{, at least in part,} in terms of \replaced[id=r23]{a }{ the} crossover from background-ion to counterion-dominated screening as the colloid concentration is increased. \added[id=r23]{The ideas are illustrated in Fig.~\ref{fgr:cartoon}.}  At low $\etac$, background salt ions dominate the screening (since ions from the dissociation of surface groups on the particles are negligible - see Fig.~\ref{fgr:cartoon}(a)) so $\kap \collrad$ is independent of the colloid concentration. The predictions of charge regulation models (which assume constant $\kap \collrad$) apply directly, and the effective charge $\Z$ reduces with increasing $\etac$. By contrast, at high packing fractions, ions \added[id=r23]{from the particles} outnumber the \added[id=r23]{bulk} salt ions  (see Fig.~\ref{fgr:cartoon}(b)). As the number of ions released from the surface of the particles depends on the number of colloids the screening parameter $\kap \collrad$ becomes $\etac$-dependent. \added[id=r23]{Electrostatic screening then \textit{increases} as the particle volume fraction grows. The increased screening at high $\etac$ reduces the coupling between charge-regulated particles, which allows the particle charge to grow.} \replaced[id=r23]{The }{Reinterpreting charge regulation predictions in terms of the average separation $\d$ between particle, we show that the} non-monotonic density dependence of $\Z$ seen in experiments is \replaced[id=r23]{therefore a consequence}{ a characteristic} of a crossover from a salt-dominated to a counterion-dominated screening regime at high $\etac$. Finally, we note that the enhanced charge repulsion at high particle concentration evident in our data may account for the unusual high colloidal stability of nanoparticle dispersions which has been reported\cite{Batista2015}.

\section*{Acknowledgments}

JEH was supported by EPSRC CDT grant EP/G036780/1 and DAJG by a studentship from Merck Chemicals. We thank Marco Heinen for his MPB-RMSA code. Finally, the authors would like to thank the ESRF and  the Diamond Light Source for X-ray beam time (experiments SC-3655 and SM8982).


\printnomenclature

%


\end{document}